\documentclass[preprint]{aastex62}
\usepackage{natbib}
\usepackage{amssymb}
\usepackage{amsmath}
\usepackage{amstext}
\usepackage{bm}
\usepackage{epsfig}
\usepackage{graphicx}
\usepackage{color}

\newcommand{\iue}{{\it IUE}}
\newcommand{\hst}{{\it HST}}

\newcommand{\niv}{N~{\sc iv}}
\newcommand{\siiv}{Si~{\sc iv}}
\newcommand{\nivlam}{N~{\sc iv}$~\lambda 1718$}
\newcommand{\siivlam}{Si~{\sc iv}$~\lambda 1402$}
\newcommand{\kms}{km s$^{-1}$}
\newcommand{\etal}{et al.}
\newcommand{\xper}{$\xi$~Per}
\newcommand{\cxo}{{\it Chandra}}
\newcommand{\xmm}{{\em XMM-Newton}}

\received{January 1, 2018}
\revised{January 7, 2018}
\accepted{\today}
\submitjournal{ApJ}

\shorttitle{X-ray and wind line variability}

\shortauthors{Massa, Oskinova, Prinja \& Ignace}
\begin{document}

\title{Coordinated UV and X-ray spectroscopic observations of the O-type 
giant \xper: the connection between X-rays and large-scale wind 
structure\footnote{Based on observations made with the NASA/ESA Hubble Space 
Telescope, obtained at the Space Telescope Science Institute, which is 
operated by the Association of Universities for Research in Astronomy, Inc., 
under NASA contract NAS 5-26555. These observations are associated with 
program \# 13760.}\footnote{Based on observations obtained with XMM-Newton, 
an ESA science mission with instruments and contributions directly funded by 
ESA Member States and NASA.}}

\correspondingauthor{Derck Massa}
\email{dmassa@spacescience.org}

\author[0000-0002-9139-2964]{Derck Massa}
\affil{Space Science Institute
4750 Walnut Street, Suite 205 
Boulder, Colorado 80301, USA} 

\author{Lida Oskinova}
\affiliation{Institute for Physics and Astronomy, University of Potsdam, 
Karl-Liebknecht-Str. 24/25, 14476 Potsdam, Germany}
\affiliation{Kazan Federal University, Kremlevskaya Str 18, 420008, Kazan, Russia}
\nocollaboration

\author{Raman Prinja}
\affiliation{Department of Physics \& Astronomy, University College London, 
Gower Street, London WC1E 6BT, UK}
\nocollaboration

\author{Richard Ignace}
\affiliation{East Tennessee State University 
Department of Physics \& Astronomy
Johnson City, TN 37614  USA}
\nocollaboration




%%%%%%%%%%%%%%%%%%%%%%%%%%%%%%%%%%%%%%%%%%%%%%%%%%%%%%%%%%%%%%%%%%%%%%%%%%%%
\begin{abstract}

We present new, contemporaneous \hst\/ STIS and \xmm\/ observations of the 
O7~III(n)((f)) star \xper.  We supplement the new data with archival \iue\/ 
spectra, to analyze the variability of the wind lines and X-ray flux of 
\xper.  The variable wind of this star is known to have a 2.086 day 
periodicity.  We use a simple, heuristic spot model which fits the low 
velocity (near surface) \iue\/ wind line variability very well, to 
demonstrate that the low velocity absorption in the new STIS spectra of 
\nivlam\/ and \siivlam\/ vary with the same 2.086 day period.  It is 
remarkable that the period and amplitude of the STIS data agree with those 
of the \iue\/ spectra obtained 22 years earlier.  We also show that the 
time variability of the new \xmm\/ fluxes are also consistent with the 2.086 
day period.  Thus, our new, multi-wavelength coordinated observations 
demonstrate that the mechanism which causes the UV wind line variability is 
also responsible for a significant fraction of the X-rays in single O stars.  
The sequence of events for the multi-wavelength light curve minima is: 
\siivlam, \nivlam, and X-ray flux, each separated by a phase of about 0.06 
relative to the 2.086 day period.  Analysis of the X-ray fluxes shows that 
they become softer as they weaken.  This is contrary to expectations if the 
variability is caused by periodic excess absorption.  Further, the high 
resolution X-ray spectra suggest that the individual emission lines 
at maximum are more strongly blue shifted.  

If we interpret the low velocity wind line light curves in terms of our 
model, it implies that there are two bright regions, i.e., regions with 
less absorption, separated by 180$^\circ$, on the surface of the star.  We 
note that the presence and persistent of two spots separated by 180$^\circ$ 
suggests that a weak dipole magnetic field is responsible for the 
variability of the UV wind line absorption and X-ray flux in \xper.
\end{abstract}


%%%%%%%%%%%%%%%%%%%%%%%%%%%%%%%%%%%%%%%%%%%%%%%%%%%%%%%%%%%%%%%%%%%%%%%%%%%%
%%%%%%%%%%%%%%%%%%%%%%%%%%%%%%%%%%%%%%%%%%%%%%%%%%%%%%%%%%%%%%%%%%%%%%%%%%%%
\section{Introduction}\label{sec:intro}
%%%%%%%%%%%%%%%%%%%%%%%%%%%%%%%%%%%%%%%%%%%%%%%%%%%%%%%%%%%%%%%%%%%%%%%%%%%%

O star $(M \gtrsim 20\,M_\odot)$ winds input significant mechanical
energy into the interstellar medium and affect the evolution of their
host clusters and galaxies.  Mass loss by winds also determine the
ultimate fate of massive stars, and the nature of their remnants.
Consequently, reliable measurements of mass loss rates due to stellar
winds are essential for all of these subjects.  Stellar winds are
driven by radiative pressure on metal lines \citep{castor75}.  Because the 
optical spectra of O stars are dominated by photospheric lines or 
recombination lines (which only sample the very base of the wind), much of 
wind research has concentrated on UV resonance lines which are formed 
throughout the wind.

Over the years, it has become apparent that radiatively driven winds are 
far more complex than the homogeneous, spherically symmetric flows 
envisioned by \citet{castor75}.  Instead, they have been shown to contain 
optically thick structures which can be quite small or very large.  Until 
we unravel the details of these flows we cannot hope to reliably translate 
observational diagnostics into physical quantities such as mass loss rates.  
To progress, we need a firm grasp on the underlying physical mechanisms 
which determine the wind structures.  The state of affairs can be seen in 
the literature, where the values of observationally derived mass loss rates 
have swung back and forth by factors of 10 or more \citep{massa03, puls06, 
fullerton06, osk07, Sundqvist11, surlan12}.  

The presence of large structures was revealed in time series of unsaturated 
UV wind lines.  Dynamic spectra of these lines display discrete absorption 
components (DACs) which indicate large, coherent structures propagating 
through the winds \citep[e.g.,][]{kaper99, prinja02}.  Similar features are 
observed in LMC and SMC O stars \citep{massa00} and in CSPNe 
\citep{prinja12}, suggesting they are a universal property of radiatively 
driven flows.  Further, \citet{prinja10} demonstrated that the wind lines 
in many OB stars have anomalous doublet ratios, indicating that the winds 
have optically thick structures embedded within them.  \citet{cranmer96} 
showed how the spiral structures produced by co-rotating interaction 
regions, CIRs \citep[e.g.,][]{mullan84}, can explain UV wind line 
variability by introducing bright spots near the equator.  The CIR model 
was applied to $\xi$~Per, O7.5~III(n)((f)) \citep{dejong01}, and HD~64760, 
B0.5~Ib \citep{fullerton97, lb08}, and its signature also appears in the 
dynamic spectra of most long duration time series \citep[e.g.,][]{prinja02, 
massa15}.  The CIR model predicts spiral structures with density 
enhancements of $\sim 2$ which, together with velocity plateaus, can 
increase Sobolev optical depths by factors of 10 to 100.  While CIRs may 
have little effect on the mass loss rate, they can strongly affect the 
observational diagnostics used to determine it.

Another indication of wind structure is the universal presence of X-ray 
emission from O stars.  Further, there is growing evidence that X-ray 
fluxes vary on time scales of days \citep{Osk2001, Naze2013, Nichols2015, 
Massa2014, rauw2015, naze2018}.  

As yet, the continuous X-ray monitoring, needed to 
determine whether this variability is periodic, is lacking.  The X-ray 
spectra of OB stars imply that the X-rays are produced throughout the wind, 
arising from the mechanical energy of impacts between different wind 
components.  It is widely accepted that the X-rays arise from the line 
de-shadowing instability, LDI \citep[e.g.,][]{lucy80, owocki88, Feldmeier97} 
In LDI models, a component of the wind flow fragments into randomly 
distributed structures which interact with each other or the ambient wind 
to produce X-rays.  Some LDI predictions agree with observations. For 
example, X-ray emission line profiles \citep[e.g.,][]{Osk2006, herve13, 
cohen14} imply that the hot plasma moves with the same velocity as the cool 
wind, and that the X-rays are attenuated by the cool gas component, as 
expected.  The doublet anomaly can also be interpreted in terms of LDI 
fragments \citep{Sundqvist11, surlan12}.  However, other predictions are at 
odds with the observations.  The model predicts strong short time-scale 
stochastic X-ray variability \citep{Feldmeier97} while none is observed 
\citep{Naze2013}.  It also predicts that the strongest shocks (hottest 
plasma) should originate well out in wind, while observations indicate that 
the hottest plasma is very close to the photosphere \citep{wc2007}.  
Further, the LDI model cannot explain X-ray variability on a time scale of 
days.  From a theoretical perspective, it is not yet clear whether LDI 
clumping and large-scale CIR structures can co-exit \citep{Sundqvist18} 
or whether LDI models can produce the observed levels X-ray fluxes when 
three dimensional effects are included \citep{stein18}.

CIRs provide an obvious, but untested, means for producing X-ray 
variability, either by creating X-rays at their interfaces with the freely 
flowing wind \citep{mullan84, ignace13} or by modulating the ambient 
X-ray flux by their density enhancements.  Two observational results that 
could enable us make progress are whether the X-ray emission varies 
periodically and whether it is related to the UV wind line variability.  If 
these are the case, it would indicate a CIR -- X-ray connection that 
could help constrain future modeling of how the CIRs and X-ray sources 
interact and their relative locations.  These tests can be performed by 
selecting a normal star that is bright in X-rays and has a well documented 
DAC period and observing in both the X-rays and UV simultaneously for more 
than one period.  These observations would determine whether the UV and 
X-ray variability are related, and they are the object of the current study.  

This paper is organized as follows: \S~\ref{sec:xiper} gives some relevant 
properties of \xper, \S~\ref{sec:obs} presents the \hst\/ and \xmm\/ 
observations, \S~\ref{sec:anal} describes our analysis, and 
\S~\ref{sec:disc} summarizes and discusses the results.

%%%%%%%%%%%%%%%%%%%%%%%%%%%%%%%%%%%%%%%%%%%%%%%%%%%%%%%%%%%%%%%%%%%%%%%%%%
%%%%%%%%%%%%%%%%%%%%%%%%%%%%%%%%%%%%%%%%%%%%%%%%%%%%%%%%%%%%%%%%%%%%%%%%%%
\section{\xper -- a rapidly rotating O type giant}\label{sec:xiper}
%%%%%%%%%%%%%%%%%%%%%%%%%%%%%%%%%%%%%%%%%%%%%%%%%%%%%%%%%%%%%%%%%%%%%%%%%%
\xper\ is a normal, single O7~III(n)((f)) star with a large, but not 
abnormal, rotational velocity, $v \sin i = 204$~\kms, and a wind terminal 
velocity $v_\infty = 2450$ \kms\ \citep{dejong01}, which is typical for 
its spectral type.  A detailed analysis of the STIS spectra will be 
presented in a forthcoming study by Hamann et al. (in prep).

\xper\ has long been a favorite star for studying wind variability.  It is 
bright and several of its UV wind lines are well developed, but 
unsaturated, making them ideal for examining variability.  Dynamic spectra 
of its UV wind lines show distinct, isolated, large amplitude, repeating 
DACs with a well-defined 2.086 day period \citep{dejong01} that persists 
for more than ten days. They assume that this period is roughly half of 
the actual rotation period and that two distinct structures, separated by 
180$^\circ$, are present in the wind.  Their assumption is bolstered by \citet{Ramiaramanantsoa14}, who obtained four weeks of contiguous, 
high-precision visual photometry of \xper.  They demonstrated that the 
observed variability was consistent with localized magnetic spots on a star 
whose rotation period is 4.18 d.  Nevertheless, it is possible that 2.086 
day could be the actual stellar rotation period (see the Appendix).

In addition to UV wind line variability, \citet{Massa2014} observed 
$\xi$~Per for 162~ks (1.88 day) with the \cxo\ HETGS.  Their results 
showed that the X-ray flux clearly varied.  However, the time baseline of 
their observations was not sufficient to determine whether the variation was 
periodic, as expected if the X-ray variability is associated with the same 
wind structures that create the DACs.

%%%%%%%%%%%%%%%%%%%%%%%%%%%%%%%%%%%%%%%%%%%%%%%%%%%%%%%%%%%%%%%%%%%%%%%%%%%%
\section{The Observations}\label{sec:obs}
%%%%%%%%%%%%%%%%%%%%%%%%%%%%%%%%%%%%%%%%%%%%%%%%%%%%%%%%%%%%%%%%%%%%%%%%%%%%
%%%%%%%%%%%%%%%%%%%%%%%%%%%%%%%%%%%%%%%%%%%%%%%%%%%%%%%%%%%%%%%%%%%%%%%%%%%%
This section describes the new observations.  We begin with the 
motivation for the time sampling employed and then discuss the reductions of 
the \hst\/ STIS and \xmm\/ data used in the analysis.  

%%%%%%%%%%%%%%%%%%%%%%%%%%%%%%%%%%%%%%%%%%%%%%%%%%%%%%%%%%%%%%%%%%%%%%%%%%%%
\subsection{Time sampling}
%%%%%%%%%%%%%%%%%%%%%%%%%%%%%%%%%%%%%%%%%%%%%%%%%%%%%%%%%%%%%%%%%%%%%%%%%%%%
Our joint \hst/\xmm\/ program requested time sampling based on the 2.086 day 
period.  We proposed to observe \xper\ 11 times with \hst\/ STIS over a time 
interval of roughly 100 hours ($\sim 5$ day) to capture a repeat of the DAC 
activity in the UV wind lines.  The \hst\/ observations were to be 
supplemented by ten 7~ks \xmm\/ observations distributed over 245 hours and 
centered on the \hst\/ observations.  These were to be obtained at 10 and 
48 hour intervals, as allowed by the \xmm\/ orbit, and sample the 2.086~day 
period over 5 periods.  

\begin{figure}
\begin{center}
  \includegraphics[width=0.75\linewidth]{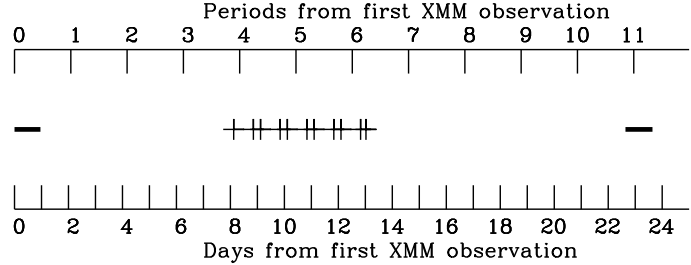}\vspace{-3.0in}
\end{center}
\vspace{-0.6in}
\caption{Observations displayed in time, with the time of the first \xmm\/ 
observation set to zero.  The lower scale is in days and the upper scale 
gives the number of 2.086 day periods.  The times covered by the \xmm\/ 
observations are shown as the solid bars and the STIS observations as 
crosses.} 
\label{fig:window} 
\end{figure}

While the \hst\/ STIS observations were performed as requested, \xmm\/ 
scheduling constraints turned out to be more severe than anticipated.
Instead of a more uniform sampling over the interval, we were allocated two 
long, (82.9 and 82.4 ks), exposures separated by 22.65 days.  Although we 
were actually given more observing time than requested, the temporal 
spacing of the observations was far from optimal.  The final time sampling 
of the observations is depicted in Figure~\ref{fig:window}.  It shows that 
the STIS observations span 2.4 periods and that each  \xmm\/ observation
spans roughly 0.46 periods and the second begins 10.52 periods after the 
first.  Consequently, although the \xmm\/ observations sample the 2.086 day 
period very well, none of them sample the same phase more than once.   

%%%%%%%%%%%%%%%%%%%%%%%%%%%%%%%%%%%%%%%%%%%%%%%%%%%%%%%%%%%%%%%%%%%%%%%%%%%%
% made by windows.pro
%%%%%%%%%%%%%%%%%%%%%%%%%%%%%%%%%%%%%%%%%%%%%%%%%%%%%%%%%%%%%%%%%%%%%%%%%%%%

%%%%%%%%%%%%%%%%%%%%%%%%%%%%%%%%%%%%%%%%%%%%%%%%%%%%%%%%%%%%%%%%%%%%%%%%%%%%
\subsection{The STIS data}
%%%%%%%%%%%%%%%%%%%%%%%%%%%%%%%%%%%%%%%%%%%%%%%%%%%%%%%%%%%%%%%%%%%%%%%%%%%%

The STIS observations used the $R=45,800$ E140M grating with the 0.2X0.05ND 
aperture, which includes a neutral density filter, necessitated because 
\xper\/ is so bright.  The spectra include the important 
\siiv$\lambda\lambda 1400$ resonance doublet and the \niv$\lambda 1718$ 
excited state line. 

Creation of the \siiv $\lambda 1402$ fluxes was relatively straight forward.  
The only complication was that the individual spectra had to be rectified 
to the band $1.0 \leq v/v_\infty \leq 1.5$.  This was necessary because the 
overall flux levels of E140M spectra obtained through the small 0.2X0.05ND 
aperture can vary due to differences in the positioning of the object in the 
aperture and variations in the image size resulting from telescope 
``breathing'' due to orbital thermal variations \citep{proffitt17}.

The extraction of the \niv\ data was more complex. Unfortunately, the 
standard STScI pipeline processing of objects observed recently by the STIS 
E140M omits the last echelle order which contains 1718 \AA\ (although 
earlier spectra include it). Consequently, we had to extract this order by 
hand.

\begin{figure}
\begin{center}
  \includegraphics[width=0.7\linewidth]{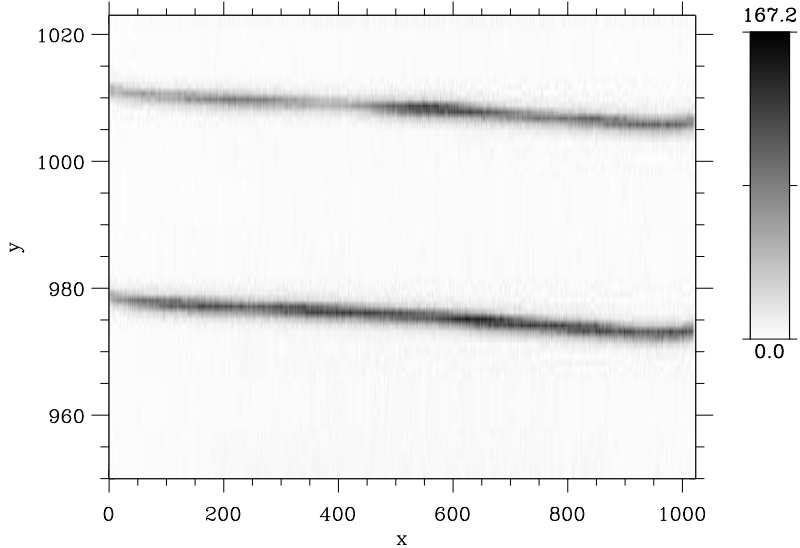}\vspace{-2.5in}
\end{center}
\caption{Image of the portion of the STIS FUV MAMA detector containing 
\niv$\lambda 1718$, the line appearing in the top order,
centered near $x = 400$.} 
\label{fig:det} 
\end{figure}

\begin{figure}
\begin{center}
  \includegraphics[width=0.7\linewidth]{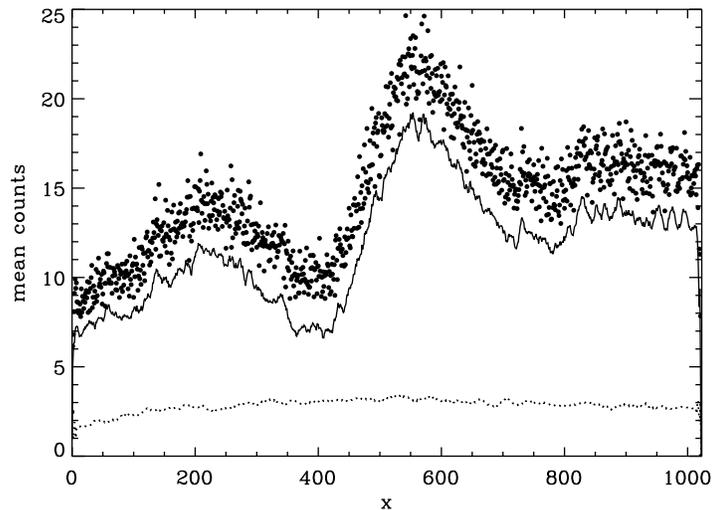}\vspace{-2.7in}
\end{center}
\caption{Raw spectrum (points), background smoothed by 16 points (dotted 
curve), and net spectrum (raw minus background) smoothed by 8 points (solid 
curve).}  
\label{fig:raw} 
\end{figure}

Figure~\ref{fig:det} shows the portion of the FUV MAMA detector which 
includes the order containing \niv$\lambda 1718$ (the upper stripe).  
Spectra for this order were extracted using a simple box car.  Specifically, 
we first determine the mean counts for $1000 \leq y \leq 1015$ as a 
function of $x$, termed the gross spectrum.  Similarly, a background 
spectrum was determined from the means over $985 \leq y \leq 1000$.  The 
background was then subtracted from the gross to give a set of un-flux 
calibrated net spectra.  Figure~\ref{fig:raw} gives an example of the 
gross, background and net spectra.   

The wavelength alignment of the net spectra appears quite good (the Helio 
centric velocities varied by less than 0.2 \kms).  The wavelength scale 
was determined by using the STIS dispersion relation applied to spectra 
which include the order containing 1718\AA.  This is $\lambda = 0.053619 x 
+ b$ \AA, where $b = 1710.75\pm 0.1$ \AA, depending on the location of the 
star and the velocity of the telescope.  An uncertainty of $\pm 0.1$ \AA\ is 
not important for our purposes.  The wavelength scale is only used to 
determine the portion of the spectrum to be averaged for studying \nivlam\/ 
flux variations.  Since an uncertainty of 0.1\AA\ only amounts to 17 \kms, 
or $\Delta v/v_\infty < 1$\%, it has an insignificant effect on the mean 
fluxes.    

As mentioned above, the individual spectra had to be rectified to the same 
mean.  In this case we used the band $800 \leq x \leq 1000$, which 
corresponds to $1753.7 \leq \lambda \leq 1764.4$, which is well away from 
expected wind activity.  Since subsequent measurements are normalized 
by their means, not absolute flux calibration is needed.

%%%%%%%%%%%%%%%%%%%%%%%%%%%%%%%%%%%%%%%%%%%%%%%%%%%%%%%%%%%%%%%%%%%%%%%%%%%%
% made by get1718.pro
%%%%%%%%%%%%%%%%%%%%%%%%%%%%%%%%%%%%%%%%%%%%%%%%%%%%%%%%%%%%%%%%%%%%%%%%%%%%
%%%%%%%%%%%%%%%%%%%%%%%%%%%%%%%%%%%%%%%%%%%%%%%%%%%%%%%%%%%%%%%%%%%%%%%%%%%%
% made by get1718.pro
%%%%%%%%%%%%%%%%%%%%%%%%%%%%%%%%%%%%%%%%%%%%%%%%%%%%%%%%%%%%%%%%%%%%%%%%%%%%

%%%%%%%%%%%%%%%%%%%%%%%%%%%%%%%%%%%%%%%%%%%%%%%%%%%%%%%%%%%%%%%%%%%%%%%%%%%%
\subsection{The \xmm\/ data}\label{sec:xmm} 
%%%%%%%%%%%%%%%%%%%%%%%%%%%%%%%%%%%%%%%%%%%%%%%%%%%%%%%%%%%%%%%%%%%%%%%%%%%%
%%%%%%%%%%%%%%%%%%%%%%%%%%%%%%%%%%%%%%%%%%%%%%%%%%%%%%%%%%%%%%%%%%%%%%%%%%%

%%%%%%%%%%%%%%%%%%%%%%%%%%%%%%%%%%%%%%%%%%%%%%%%%%%%%%%%%%%%%%%%%%%%%%%%%%%
\begin{figure}
\centering
\includegraphics[height=0.7\columnwidth]{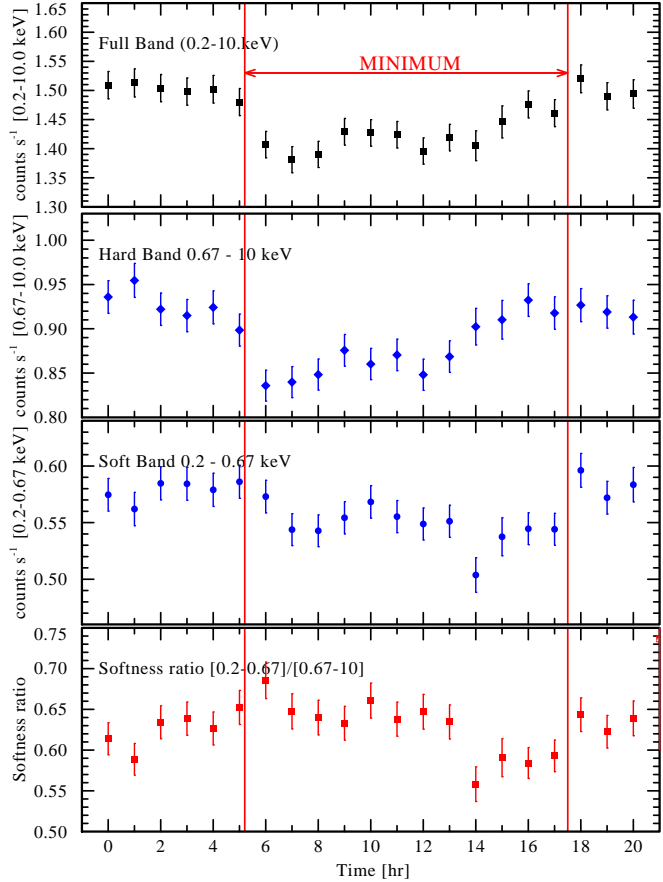}
\caption{EPIC PN background subtracted X-ray light curves of the first 
\xper\ observation.  The data were binned to 1\,hr (3.6\,ks).  The 
horizontal axis denotes the time after the beginning of the observation in 
hours.  The vertical axes in the three upper panels show the count rate as 
measured by the EPIC PN camera. The  error bars (1$\sigma$) correspond to 
the combination of the error in the source counts and the background 
counts.  The three upper panels show the light curves in different bands, 
as indicated. The lowest panel shows the ``softness ratio'' obtained by 
dividing count rate in 0.2-0.67\,keV band by the count rate in 
0.67-10.0\,keV band.  The vertical red lines encompass the time interval we 
define as the X-ray minimum.}
\label{fig:lc1}
\end{figure}
%%%%%%%%%%%%%%%%%%%%%%%%%%%%%%%%%%%%%%%%%%%%%%%%%%%%%%%%%%%%%%%%%%%%%%%%%%%
%%%%%%%%%%%%%%%%%%%%%%%%%%%%%%%%%%%%%%%%%%%%%%%%%%%%%%%%%%%%%%%%%%%%%%%%%%%
\begin{figure}
\centering
\includegraphics[height=0.7\columnwidth]{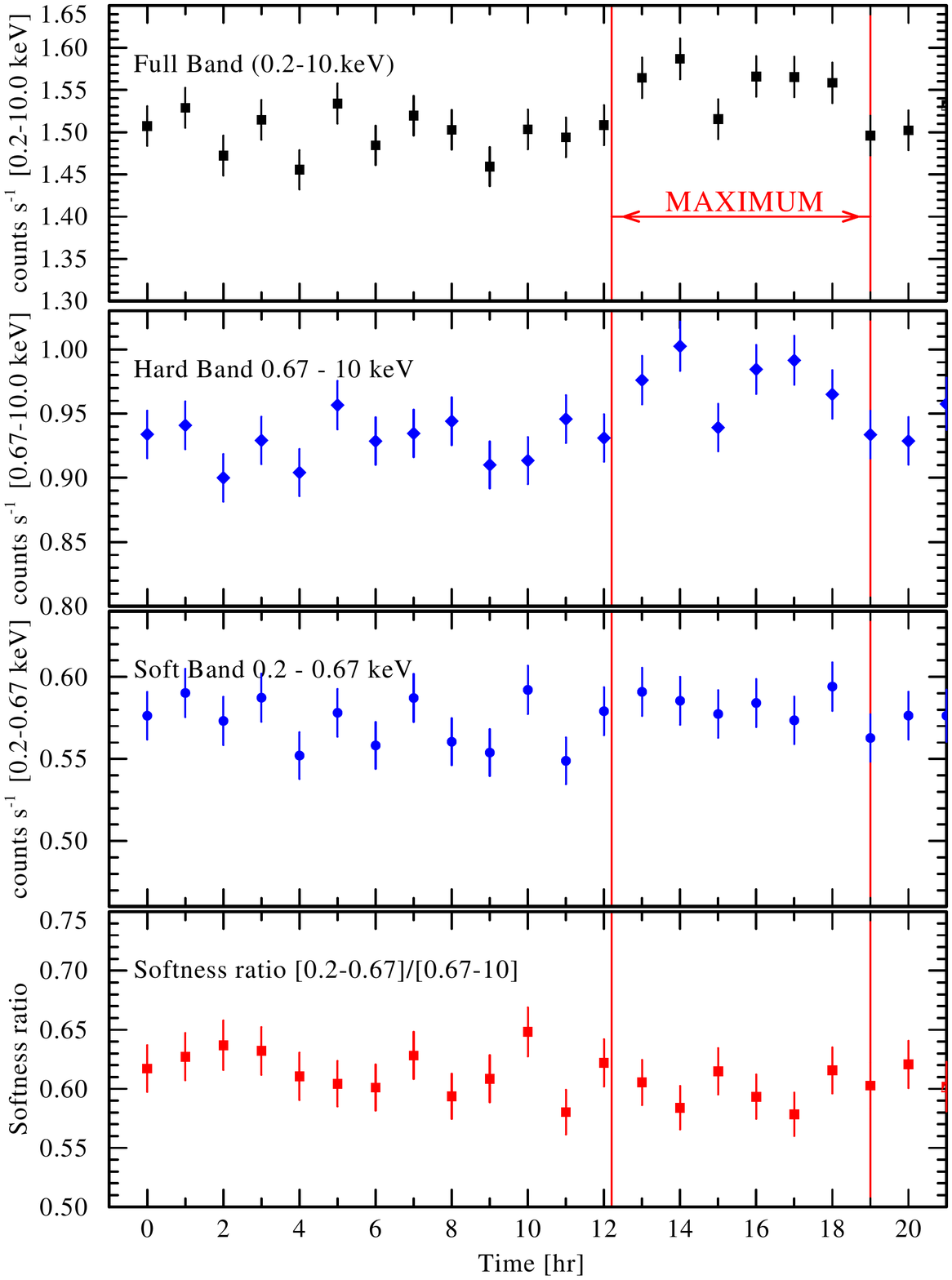}
\caption{The same as in Fig.\,\ref{fig:lc1}, but for the second observation. 
The vertical red lines encompass the time interval we define as maximum.}
\label{fig:lc2}
\end{figure}
%%%%%%%%%%%%%%%%%%%%%%%%%%%%%%%%%%%%%%%%%%%%%%%%%%%%%%%%%%%%%%%%%%%%%%%%%%%

\xmm\/ observed \xper\/ on two separate visits, first for a 82.9\,ks exposure 
on 2016-02-06 (ObsID 0770990101), and again for a 82.4\,ks exposure on 
2016-02-28 (ObsID 0770990201) (see Fig. \ref{fig:window}).  The three \xmm\/ 
X-ray telescopes illuminate five instruments which operate simultaneously 
and independently: two reflection grating spectrometers \citep[RGS1 and 
RGS2,][]{rgs2001}, with a spectral resolution of $\sim 0.07$\,\AA\ and 
wavelength coverage of $5 \lesssim \lambda \lesssim 38$\,\AA, and three 
focal plane instruments, forming the European Photon Imaging camera (EPIC).  
The EPIC camera consists of the MOS1 and MOS2 (Metal-Oxide Semiconductor) 
and PN (pn-CCDs) detectors.  The EPIC instruments cover $1.2 \lesssim 
\lambda \lesssim 60$\/\AA, with a spectral resolution of ($E/\Delta 
E\approx 20 - 50$) \citep{mos2001,pn2001}.  During each observation, all 
three EPIC cameras were operated in the standard, full-frame mode with a 
thick UV filter and simultaneous RGS1 and RGS2 spectra were acquired. The 
optical monitor was not operating due to the excessive optical brightness 
of \xper. The data were reduced using the most recent calibrations and the 
Science Analysis System (SAS) v.15.0.  The spectra and light-curves were 
extracted using standard procedures.  The background area was chosen to be 
nearby the star and free of X-ray sources.  

%%%%%%%%%%%%%%% Light curves from the EPIC spectra
X-ray light curves were constructed from the EPIC data to examine the 
temporal variability.  The EPIC data were binned into  1\,hr (3.6\,ks) time 
bins and three broadband wavelength bins:  the ``full'' (0.2-10.0\,keV)  
EPIC bandpass, the ``soft'' (0.2-0.67\,keV) band and the ``hard'' (0.67-
10.0\,keV) band.  The X-ray light curves for these bands measured by the 
EPIC PN camera, as well as a ``softness'' ratio of the soft and hard bands 
are shown in Figures \ref{fig:lc1} and \ref{fig:lc2} for the first and 
second observations, respectively.  The light curves measured by the EPIC 
MOS cameras are similar. All of the X-ray light curves show statistically 
significant variability of $\gtrsim 10$\%\ on a time scale of hours.  
As pointed out in \S~\ref{sec:intro}, similar variability has been detected 
in all other sufficiently well studied O-stars. 

The full band light curves in Figs.\,\ref{fig:lc1} and  \ref{fig:lc2} 
contain time intervals when the count rate is lower and higher than the 
average.  We denote these time intervals as ``minimum'' and ``maximum''. 
The minimum occurred during first observation and the maximum during the 
second observation. The lowest panels in Figs.\,\ref{fig:lc1} and 
\ref{fig:lc2} show the ratio of counts rates in soft and hard bands, termed 
the ``softness ratio''. 

\begin{figure}
\centering  
\includegraphics[width=0.5\columnwidth, angle =-90]{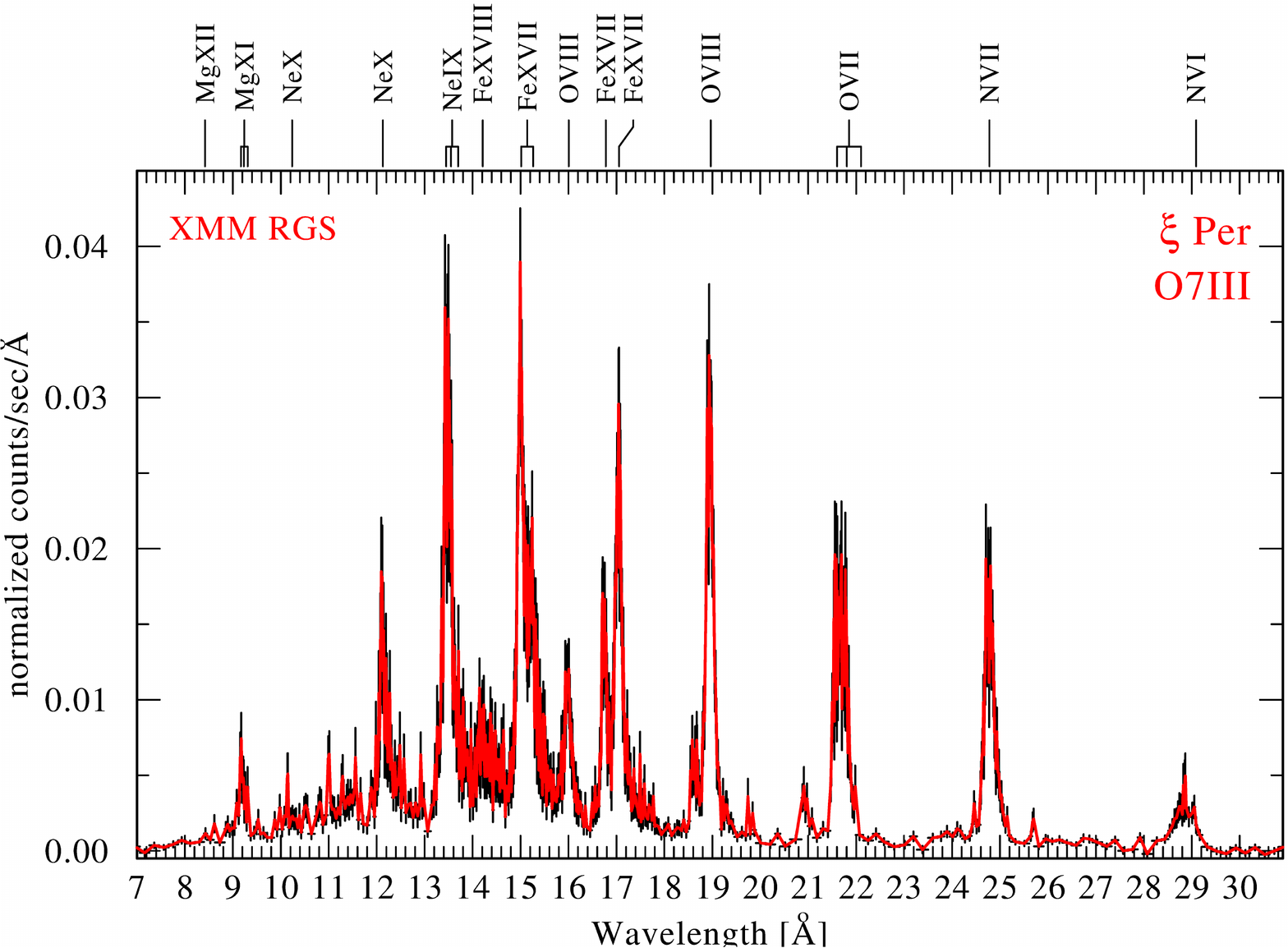} 
\caption{\xmm\/ combined RGS1+2 spectrum of \xper\/ integrated over the 
full exposure time (red curve).  Strong emission lines are identified.
The error bars (black) correspond to 3 $\sigma$.}
\label{fig:rgscomb}
\end{figure}

%%%%%%%%%%%%%%%%%%%% RGS spectra 
Finally, Figure\,\ref{fig:rgscomb} shows the merged RGS spectrum of \xper.  
Overall, its mean spectrum is not unusual for an O star of its temperature 
 \citep{wc2007, walborn09, cohen14}.  

%=============================================================
%==============================================================

%\clearpage

%%%%%%%%%%%%%%%%%%%%%%%%%%%%%%%%%%%%%%%%%%%%%%%%%%%%%%%%%%%%%%%%%%%%%%%%%%%
%%%%%%%%%%%%%%%%%%%%%%%%%%%%%%%%%%%%%%%%%%%%%%%%%%%%%%%%%%%%%%%%%%%%%%%%%%%
\section{Analysis}\label{sec:anal}
%%%%%%%%%%%%%%%%%%%%%%%%%%%%%%%%%%%%%%%%%%%%%%%%%%%%%%%%%%%%%%%%%%%%%%%%%%%

\subsection{Variable UV Wind Lines}\label{sub:UV}

Our ultimate goal is find a link between the variability seen in the UV 
wind lines and the X-rays.  This would imply that whatever causes the UV 
DACs is also responsible for the X-ray variability.  But first, we must 
demonstrate that the wind activity during our observations was similar to 
that expected from the earlier \iue\/ time series.  Because the new data 
are temporally scattered and sparsely sample the phase, we appeal to 
previous \iue\/ observations for guidance.  Consequently, we begin the 
analysis with a reexamination of existing \iue\/ observations.  First, we 
explain why we elect to analyze specific lines over a limited range in 
velocity.  Next we introduce a simple model which captures the variations 
seen in the \iue\/ data.  We then use the model to demonstrate that the 
new STIS data are consistent with the previous observations and to 
determine the relative phases of the UV lines and the \xmm\/ light curve.  
Finally, we examine how the \xmm\/ spectra respond to changes in the X-ray 
intensity.  

%%%%%%%%%%%%%%%%  Line selection 
The available UV spectral time series include two wind lines that are well 
developed but unsaturated, the property required to study variations.  
These are the \siiv$\lambda\lambda 1400$ resonance doublet and the 
\niv$\lambda 1718$ excited state line.  We would like to use these UV 
lines to extract information about the source of the DACs.  This means 
that we want to sample the wind as close to the source as possible, 
at low wind velocity.  At high velocity the modulation of the flux can 
become complex due to the evolution of the structures responsible for the 
DACs as they move through the turbulent wind and to the overlap of 
different structures in velocity \citep[e.g.,][]{puls93, cranmer96}.

The \nivlam\/ excited state line is of particular importance.  The lower 
level of this line depends on the stellar radiation field to populate it 
\citep[][]{olson81}.  Therefore, we can be certain that it samples the 
wind very close to the star \citep{massa15}.  Extracting meaningful 
information from the \siiv~$\lambda\lambda 1400$ doublet at low velocity is 
problematic because the separation of the components is $w=0.8$, where $w = 
v/v_\infty$, $v = c(\lambda -\lambda_0)/\lambda$, and $\lambda_0$ is the 
rest wavelength of the line.  This means that the high speed wind 
absorption from the red component affects the absorption by the blue 
component between $-0.2 \leq w \leq 0.0$.  As a result, for this line it is 
not possible to extract an uncontaminated measure of the wind activity close 
to the star.  In contrast, all of the absorption by the red component is 
affected by emission from the blue component.  However, it is well known 
that, for P-Cygni lines, the emission originates from throughout the wind 
and tends to be far less variable than the blue absorption, which originates 
from a column between the observer and the stellar disk 
\citep[e.g.,][]{mega, kaper96}. Consequently, one can expect the low speed 
absorption of the 1402\AA\ component to be relatively free of variable 
influences due to the weak emission from the 1393\AA\ component.  Therefore, 
we include data from this component in our analysis.

\begin{figure}
\begin{center}
  \includegraphics[width=0.45\linewidth]{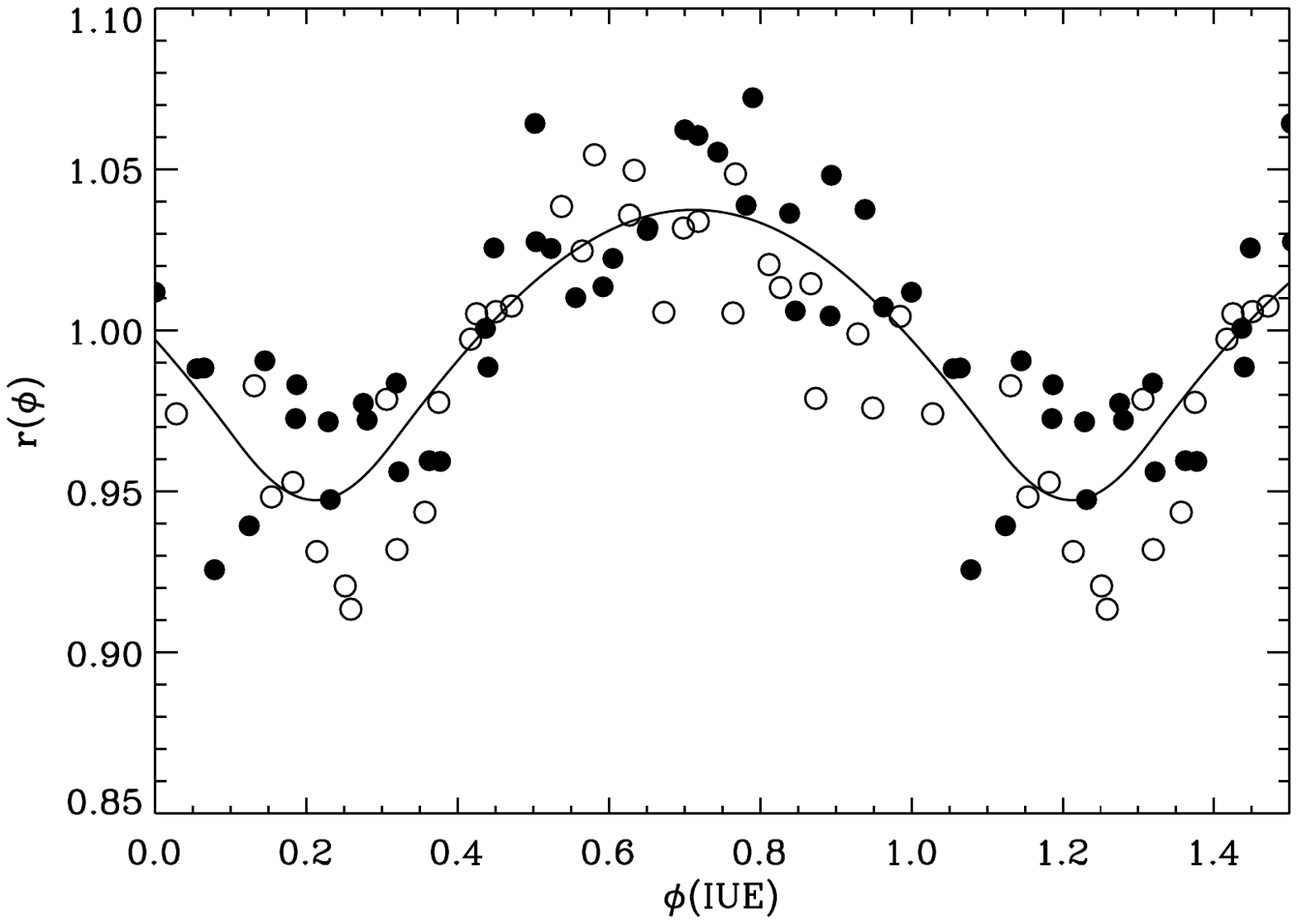}\vspace{-1.0in}
  \includegraphics[width=0.45\linewidth]{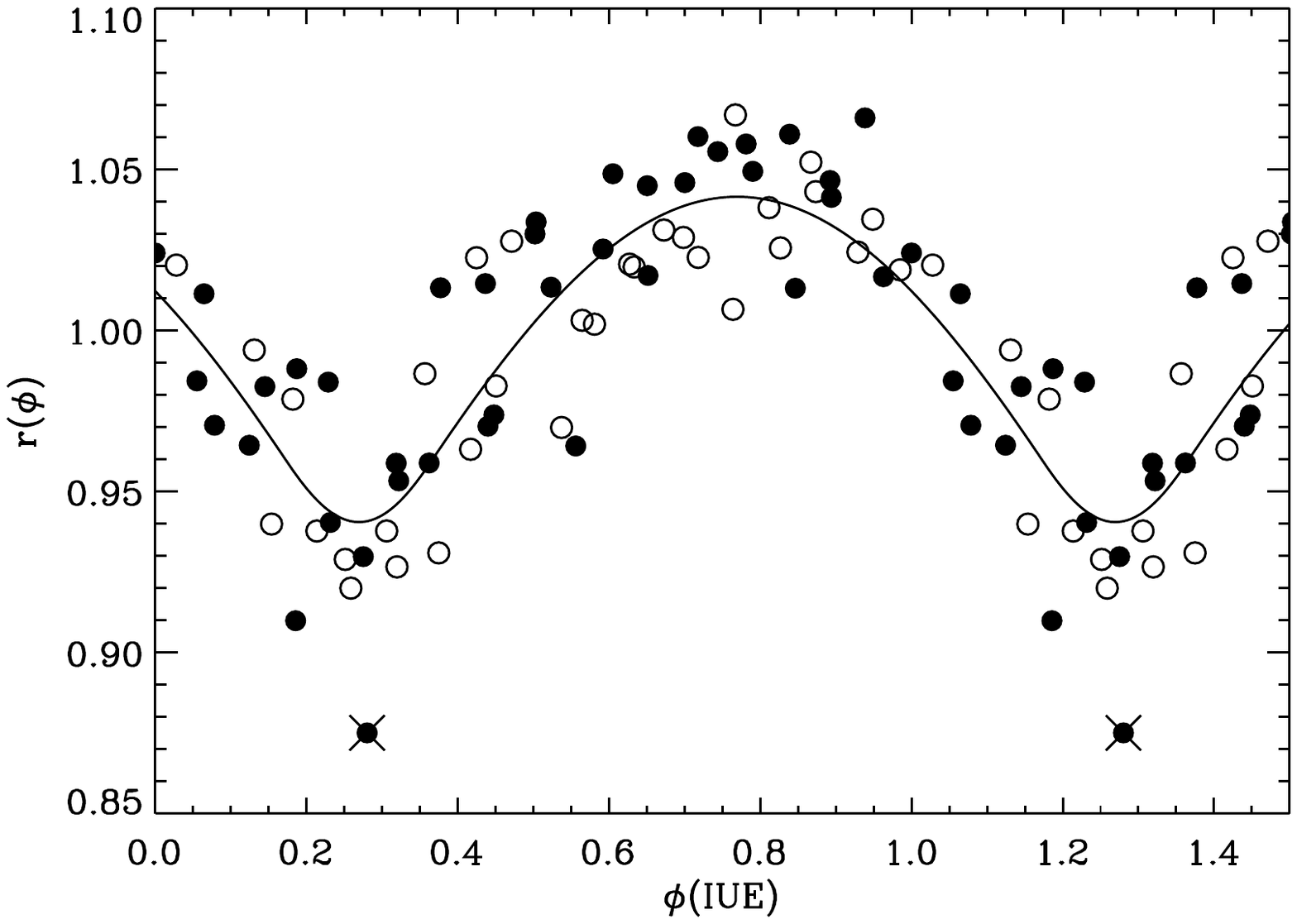}\vspace{-0.7in}
\end{center}
\caption{Normalized \iue\/ wind line fluxes for \xper\ binned over 
$-0.2 \leq v/v_\infty \leq 0$ for \siiv$\lambda 1402$ (left) and \niv 
$\lambda 1718$ (right) plotted against phase, $\phi$, relative to the 
2.086~day period.  Filled circles are for odd phases and open circles are 
for even phases, which are thought to arise from different spots.  The solid 
curves are fits to the data using eq.~(\ref{eq:model}), resulting in the 
parameters listed in Table~\ref{tab:params} .  The point in the \niv\/ 
plot with a cross through it is discordant with its neighbors and was given 
zero weight in the fit.  \label{fig:iue_fits} }
\end{figure}

%%%%%%%%%%%%%%%%%% IUE data 
To characterize the variations, we employ the well studied \iue\/ time 
series of \xper\ obtained in October 1994 \citep[e.g.,][]{dejong01, 
massa15}.  We begin by normalizing the spectra over the range  $1.0 
\leq w \leq 1.5$.  This accounts for uncertainties in the absolute flux 
levels of \iue\/ high dispersion spectra.  Figure~\ref{fig:iue_fits} shows 
the flux variation in \niv$\lambda 1718$ averaged over the region $-0.2 \leq 
w \leq 0.0$, where the fluxes are normalized by their mean value and plotted 
against phase, relative to the 2.086~day period determined by deJong \etal\/  
Points from even and odd cycles are plotted with different symbols to 
examine whether the variations caused by the presumably two distinct spots 
differ.  There does not appear to be a discernible difference.  The flux in 
the line varies by about 10\%.  The \siiv $\lambda1402$ fluxes were 
similarly binned and normalized to obtain the light curve shown in 
Figure~\ref{fig:iue_fits}.   It is clear that the shape and amplitude of 
the flux variations in \niv\ and \siiv\ are very similar.  

%xxxxxxxx
Echelle data are subject to systematic errors which arise from 
placement of the object in the aperture and instantaneous telescope 
focus.  Because these can be much larger than the statistical errors, we 
used a direct method to estimate the errors.  First, a nearby continuum 
location was selected and the same number of wavelength points used to bin 
the line data between $-0.2 \leq v/v_\infty \leq 0.0$ were binned in the 
normalized continuum region of each spectrum.  Next, we calculated the 
sample variance of the binned points.  The results were $\gtrsim 0.01$.  
Since the continuum points are divided by a larger mean flux than in the 
lines, the errors in the line data are expected to be of order 0.02.

%%%%%%%%%%%%%%%%% Enter the model
We would like to codify the shapes of the light curves shown in 
Figure~\ref{fig:iue_fits} by a function which can faithfully represent them 
and also provide some insight into the structure of the wind.  Such a 
function will also be useful for analyzing the sparsely sampled STIS data 
described below.  To do this, we developed a simple model consisting of a 
wind whose strength differs from the global flow only in a pair of 
identical, uniform, equatorial, circular spots separated by 180$^\circ$. 
Our simple model is for two diametrically opposed spots on the equator of 
the star.  In fact, if $\sin i \sim 1$, these diametrically opposed spots 
could also be at higher latitude.  Similarly, if the rotation period of 
\xper\ is 2.068 day and $i \simeq 45^\circ$ (see the Appendix), the light 
curve could be due to a single, high latitude spot.  Regardless, in all 
cases, the major factors that determine the shapes of the model light curves 
are the size of the spots and the fraction of the cycle that they are 
occulted by the star.  As we shall see, the simplest case of two equatorial 
spots provides an adequate fit to the data.  Consequently, there is little 
reason for adding additional parameters to the model since they would be 
poorly determined.  Nevertheless, we emphasize that the variability could be 
due to spots at higher latitude or even due to a single, high latitude spot 
if the rotation period is 2.086 day.

Our two spot model is described in the Appendix, where it is shown that it 
results in a normalized light curve, $r(\phi_i)$, of the form 
\begin{equation}
r(\phi_i) = \frac{1 +C a(a_0, \phi_i +\phi_0)}{1 +C \langle a(a_0, \phi 
            +\phi_0)\rangle}  \;\; .  \label{eq:model}
\end{equation}
In this expression, $\phi_i$ is the phase of the $i$-th measurement 
relative to the 2.086~day period, $C$ is a constant, $a(a_0, \phi_i+\phi_0)$ 
is the fraction of the star covered at $\phi_i$, $a_0 = \pi R_{spot}^2/
(\pi R_\star^2)$ is the fractional area of the spot at $\phi = 0$, and 
$\phi_0$ is the phase shift needed to align the model curve with the 
observations.  The shape of $r(\phi)$ depends on the three parameters: $C$, 
$a_0$ and $\phi_0$.  Some properties of the function are discussed in the 
Appendix.  Fits to the \iue\/ data using this function are shown in 
Figure~\ref{fig:iue_fits}.  The fits were determined by unweighted, 
non-linear least squares, using the Interactive Data Language (IDL) 
procedure MPFIT developed by C.\ Markwardt\footnote{The Markwardt IDL 
Library is available at {http://cow.physics.wisc.edu/$\sim$craigm/idl/}.}.  
The parameters derived from the fits are listed in Table~\ref{tab:params}, 
along with the RMS residuals of the fits, which are comparable to the 
observational errors.  Note that there is a significant phase difference 
between the \siiv\ and \niv\ light curves, $\phi_0($\siiv$) -\phi_0($\niv$) 
= 0.057 \pm 0.016$. 

A caution is in order concerning the mono-variate errors listed in 
Table~\ref{tab:params}.  The error quoted for $\phi_0$ is quite robust, 
because the full error covariance matrix of the parameter errors shows that 
it is independent of the other variables, with correlation coefficients all 
less than 0.15.  In contrast, the parameters $C$ and $a_0$ are highly 
correlated, for reasons given in the appendix.  The correlation coefficients 
for these two parameters are $-0.89$ and $-0.91$  for the \niv\ and \siiv\ 
fits, respectively.  This means that even though the derived values have 
small mono-variate errors, their values can be changed substantially and 
have little effect on the quality of the fits, as long as $a_0 C$ is held 
constant. 
%%%%%%%%%%%%%%%%%%%%%%%%%%%%%%%%%%%%%%%%%%%%%%%%%%%%%%%%%%%%%%%%%%%%%%%%%%%%
% made by iue_wind_mp3.pro
%%%%%%%%%%%%%%%%%%%%%%%%%%%%%%%%%%%%%%%%%%%%%%%%%%%%%%%%%%%%%%%%%%%%%%%%%%%%

%%%%%%%%%%%%%%%%%%%%%%%%%%%%%%%%%%%%%%%%%%%%%%%%%%%%%%%%%%%%%%%%%%%%%%%%%%%%
\begin{table}[htp]
\caption{\iue\/ Model Parameters}
\begin{center}
\begin{tabular}{lrrrr} \hline
Line & \multicolumn{1}{c}{$\phi_0$} & \multicolumn{1}{c}{$a_0$} &  
\multicolumn{1}{c}{$C$} & \multicolumn{1}{c}{$RMS$} \\ \hline
\iue\ 1402 & $-0.288 \pm 0.012$ & $0.112 \pm 0.021$ & $1.02 \pm 0.20$ & 0.022 \\
\iue\ 1718 & $-0.231 \pm 0.011$ & $0.094 \pm 0.013$ & $1.34 \pm 0.21$ & 0.022 \\ 
STIS 1402  & $-0.531 \pm 0.025$ & $0.100 \pm 0.000$ & $1.17 \pm 0.40$ & 0.021 \\
STIS 1718  & $-0.414 \pm 0.042$ & $0.100 \pm 0.000$ & $0.97 \pm 0.37$ & 0.021 \\ 
\xmm\      & $-0.335 \pm 0.020$ & \multicolumn{1}{c}{$\cdots$} &  
\multicolumn{1}{c}{$\cdots$} &  \multicolumn{1}{c}{$\cdots$} \\ \hline
\end{tabular}
\end{center}
\label{tab:params} 
\end{table}

% latest from iue_wind_mp3.pro with one N IV pt ignored
%********** 1718 **************************
%phi0  =  -0.23051 +/- 0.0107
%a     =   0.09400 +/- 0.0134
%C     =   1.34166 +/- 0.2143
%chi   =   1.10861
%RMS   =   0.02217
%tau0 >=   0.85086 +/- 0.0915
%********** 1400 **************************
%phi0  =  -0.28797 +/- 0.0123
%a     =   0.11174 +/- 0.0208
%C     =   1.01513 +/- 0.1976
%chi   =   1.08437
%RMS   =   0.02169
%tau0 >=   0.70068 +/- 0.0980

%
%STIS 1400
%phi0 = -0.535+/-0.025, a0 = 0.100+/-0.000, C = 1.170+/-0.404, RMS = 0.02147
%STIS 1718 
%phi0 = -0.414+/-0.042, a0 = 0.100+/-0.000, C = 0.972+/-0.336, RMS = 0.02115

%IUE   phi(1718) -phi(1402)  = 0.057 +/- 0.016
%STIS  phi(1718) -phi(1402)  = 0.122 +/- 0.049
%delta dphi(STIS) -dphi(IUE) = 0.065 +/- 0.052 = 1.25 sigma
%print, 0.122 -0.057, '+/-', sqrt(0.016^2+0.049^2) = 0.0650+/-0.0515
%
%        corr14                        corr17
% 1.00000 -0.90795 -0.15561   1.00000 -0.88651  0.07261
%-0.90795  1.00000  0.13842  -0.88651  1.00000 -0.08407
%-0.15561  0.13842  1.00000   0.07262 -0.08407  1.00000

%%%%%%%%%%%%%%%%%%%%% Analysis of the light curves 

\begin{figure}
\begin{center}
  \includegraphics[width=0.45\linewidth]{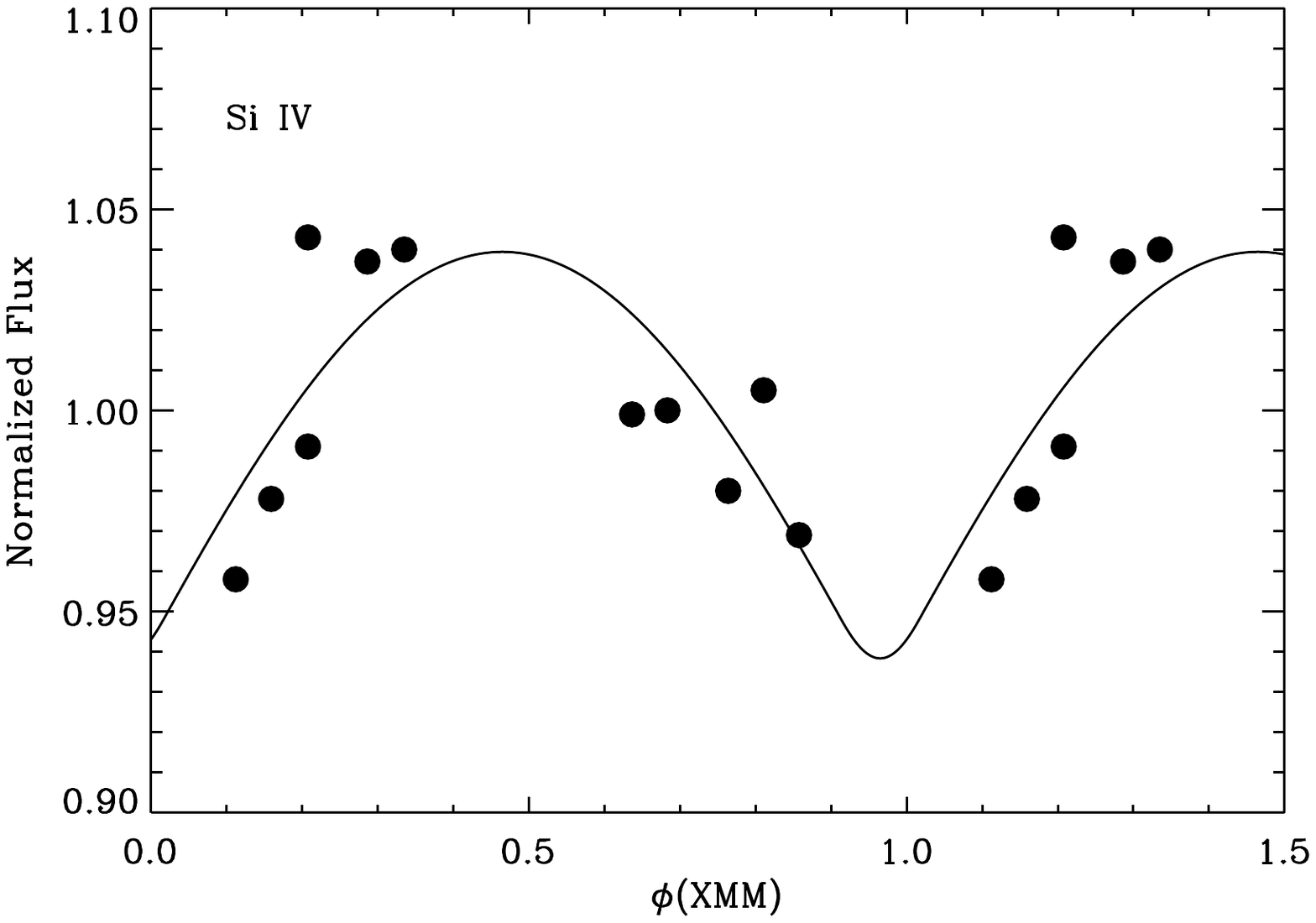}\hfill % \vspace{-0.80in}
  \includegraphics[width=0.45\linewidth]{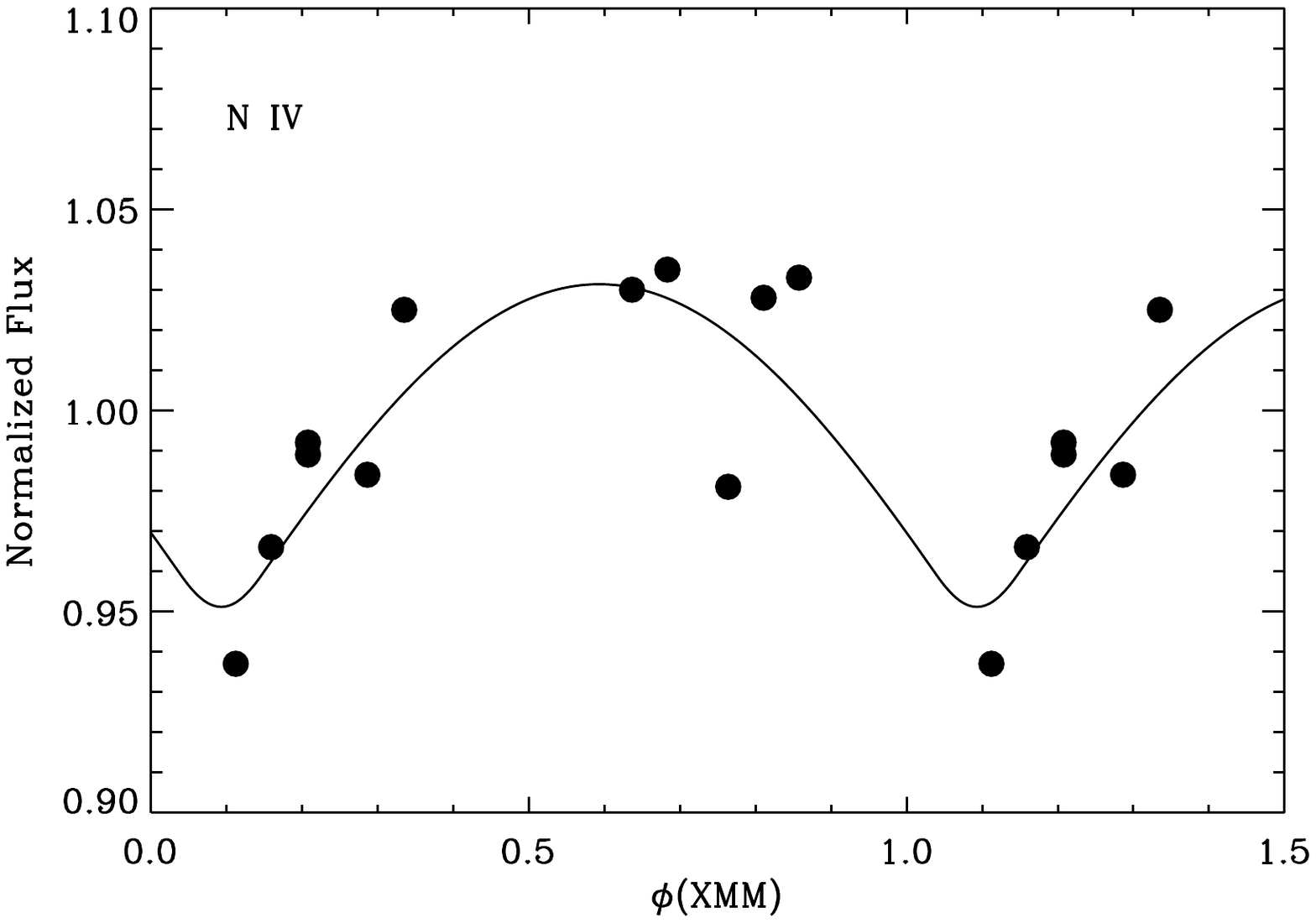}\vspace{-1.7in}
\end{center}
\caption{Model fit to the normalized STIS light curves for \siiv$\lambda 
1402$ (left) and \niv$\lambda 1718$ (right).  The parameters for the fits 
are given in Table~\ref{tab:params}.  The phases are relative to the first 
\xmm\/ observation.}  
\label{fig:stis_fits} 
\end{figure}

A set of STIS light curves were prepared in the same way as the \iue\/ data.  
Figure~\ref{fig:stis_fits} shows the STIS fluxes averaged over $-0.2 \leq w 
\leq 0.0$, for both \niv $\lambda 1718$ and \siiv $\lambda 1402$.  Both 
curves are normalized by their mean.  Observational errors were determined 
as before and are also expected to be about 0.02 in the lines.

Although the parameters $a_0$ and $C$ are highly correlated, the \iue\/ data 
are able to determine them reasonably well because the curves are so well 
sampled.  This is not the case for the STIS data.  To constrain the fits to 
the STIS data, $a_0$ was fixed at 0.1, which is similar to the \iue\/ values.  
This does not degrade the quality of the fits compared to those with both 
parameters free, and it does not alter the value of $\phi_0$.  Its only 
effect is that the values of $C$ determined by the fits are closer to the 
\iue\/ results.  Without the constraint, very different values of $a_0$ and 
$C$ result, but their product, $a_0 C$, is nearly identical to that of the 
parameters listed in Table~\ref{tab:params}.  The important point is that 
all reasonable fits result in $a_0 C >0$.  As described in the appendix, 
this implies that the spots are brighter than their surroundings.  

The fits to the STIS data are displayed in Figure~\ref{fig:stis_fits} and 
their parameters are listed in Table~\ref{tab:params}.  Once again, the fits 
are considered excellent, with RMS residuals of $\simeq 2$\%.  
The relative phase between the \niv\ and \siiv\ fits determined by the STIS 
curves is $0.127 \pm 0.049$.  As with the \iue\/ data, the shift appears to 
be real.  Although the STIS difference is larger than \iue, the difference 
between the two is only slightly larger than $1 \sigma$, i.e., $0.070 \pm 
0.052$, or $1.4 \sigma$.  Thus, both the STIS and \iue\/ data yield 
consistent amplitudes and relative phases.  It is remarkable that data 
taken 21 years apart appear so similar.

The values in Table \ref{tab:params} show that $C$ is $\sim 1$\ for the 
\iue\ data.  We employ equation (\ref{eq:const}) in the Appendix to examine 
the implications of this value.  First consider the case where the bright 
spot results from a reduction in the photospheric absorption line in the 
area occupied by the spot.  To obtain a $C \simeq 1$, requires $f_s/f_0 
\simeq f_e/f_0 +2$, or a weakening of the line flux by more than a factor of 
2.  Next, consider the case where the bright spot is a region that is free 
of low speed wind absorption and the surrounding wind has an optical depth 
of $\tau_w$.  In this case, the same equation becomes $(f_s -f_s 
e^{-\tau_w})/(f_s e^{-\tau_w} +f_e)$.  If $\tau_w \simeq 1$, then for $C = 
1$, $f_e/f_s \simeq 0.25$.  Both results are reasonable.  

There are a few properties of the fits worth emphasizing.  First, the RMS 
residuals are $\sim 2$\%, which is similar to the expected errors.  This is 
somewhat surprising, considering that the observations were obtained over 
several cycles and are possibly due to two distinct spots.  Second, both 
sets of data imply that there is a significant phase difference between 
the \siiv\ and \niv\ light curves.  Third, within the context of our simple 
model, the observed light curves can only be fit by curves with $C > 0$, 
which implies the spots are brighter than their surroundings (see Appendix).  

\subsection{Variable X-ray Emission}\label{sub:xray}

%%%%%%%%%%%%%%%%%%%%%%%%%%%%%%%%%%%%%%%%%%%%%%%%%%%%%%%%%%%%%%%%%%%%%%%%%%%%
% made by iue_wind_mp3x.pro
%%%%%%%%%%%%%%%%%%%%%%%%%%%%%%%%%%%%%%%%%%%%%%%%%%%%%%%%%%%%%%%%%%%%%%%%%%%%

%%%%%%%%%%%%%%%%%%%%%%%%%%%%%%%%%%%%%%%%%%%%%%%%%%%%%%%%%%%%%%%%%%%%%%%%%%%%
% made by new_xmm.pro and addendumx.pro
\begin{figure}[!htb]
\centering
  \begin{minipage}{0.45\textwidth}
    \centering
    \includegraphics[width=1.\linewidth, height=0.5\textheight]{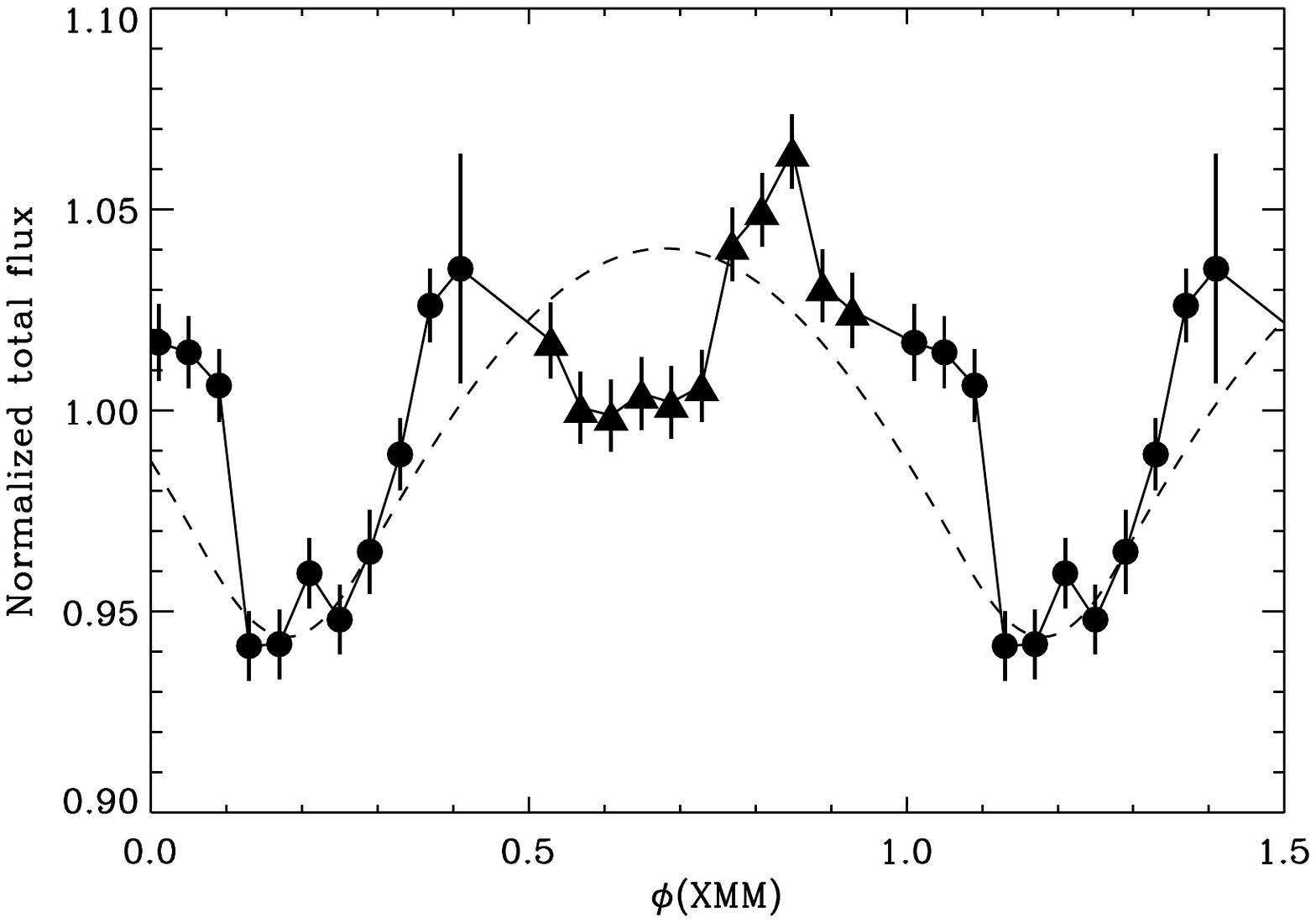}
    \vspace{-2.3in}
    \caption{Normalized sum of the \xmm\/ PN, MOS1 and MOS2 data and their 
    $1 \sigma$ error bars along with the \iue\/ \niv\ model (dashed curve) 
    shifted to align with the minima.  The phases are relative to the first 
    \xmm\/ observation.  Circles and triangles represent data from the first 
    and second set of observations.}
    \label{fig:xmm_mean} 
  \end{minipage}\hspace{0.15in}
  \begin{minipage}{0.45\textwidth}
    \centering
    \includegraphics[width=1.\linewidth, height=0.5\textheight]{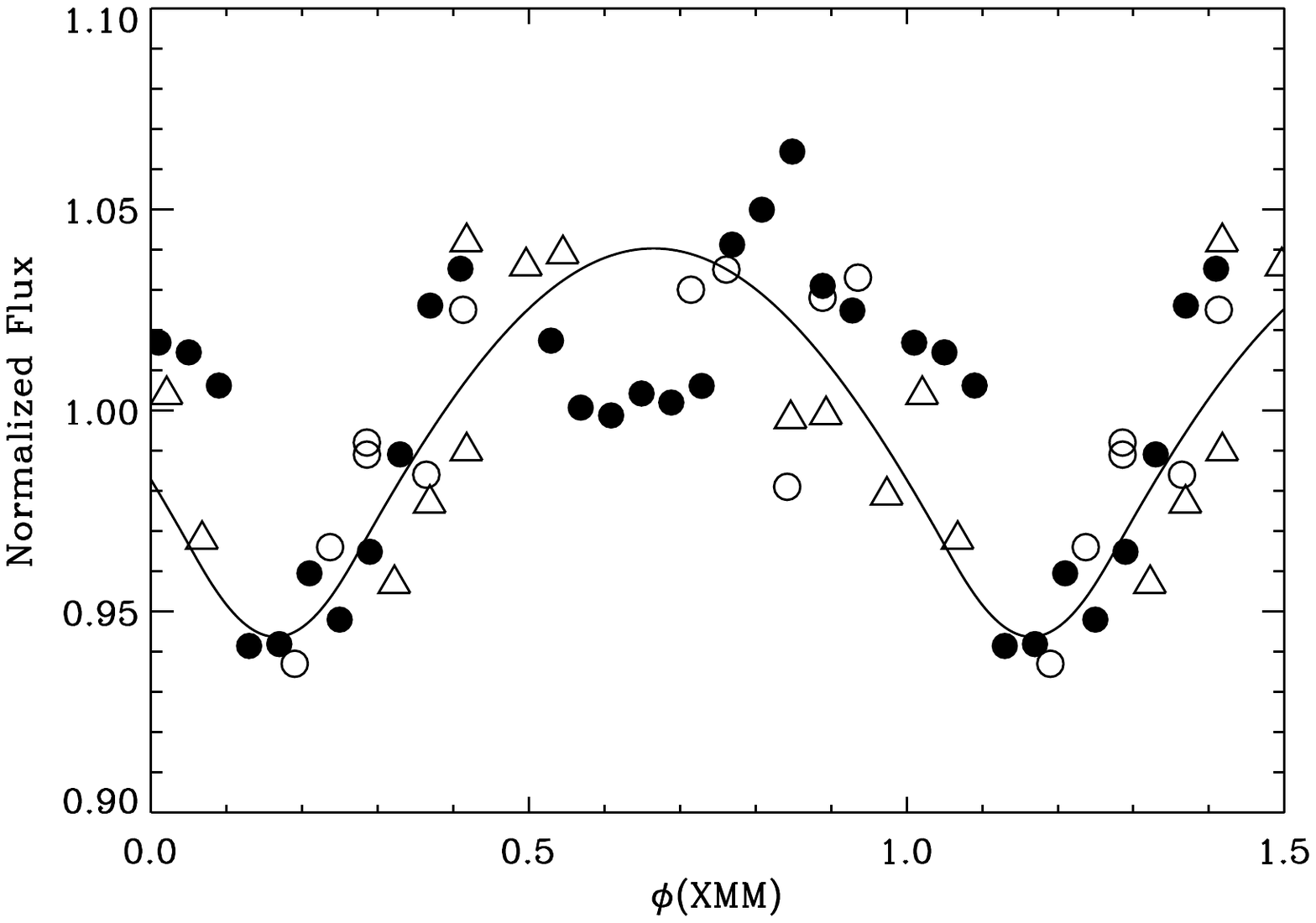}
    \vspace{-2.3in}
    \caption{Normalized \xmm\/ and STIS data aligned.  Filled circles are 
    for the \xmm\/ fluxes, open circles for the \niv\ data, and open 
    triangles for the \siiv\ data.  The STIS data have been shifted to 
    align them with the \xmm\/ data to emphasize how they all appear to 
    follow a single period.} 
    \label{fig:all_aligned}
  \end{minipage}
\end{figure}
%%%%%%%%%%%%%%%%%%%%%%%%%%%%%%%%%%%%%%%%%%%%%%%%%%%%%%%%%%%%%%%%%%%%%%%%%%%%

%%%%%%%%%%%%%%%%%%%%%%% The XMM fluxes
The \xmm\/ data were obtained over a 23 day interval (Fig.~\ref{fig:window}) 
with a 21 day gap between two roughly day long exposures.  Since the 
expected period is 2.086~day, the two observations are separated by about 10.5 
cycles.  This makes aligning the data very sensitive to the period assumed.  
Figure~\ref{fig:xmm_mean} shows the normalized \xmm\/ light curve with 
different symbols indicating when the observations were obtained.  The curve 
shape is distinctly different from those of the lines, with a broad shoulder 
after the minimum and a sharp peak before it.  The amplitude of the 
variation is roughly 10\%.  We also show a model light curve.  Since the 
\xmm\/ light curve does not resemble those of the wind lines, we simply 
shifted the best model fit to the \iue\/ \nivlam\/ data until it agreed with 
the flux minima.  Our best estimate for the phase shift and its error are 
listed in Table~\ref{tab:params}.  The error was also estimated by 
emphasizing when the minimum was clearly not aligned with the model.  It is 
unfortunate that the observations do not overlap in phase, although the two 
segments of the light curve do appear to join smoothly.  

Figure~\ref{fig:all_aligned} shows the STIS data shifted to align with the 
\xmm\/ fluxes to emphasize their mutual phase dependence.  Using the 
$\phi_0$ values listed in Table~\ref{tab:params}, we determine the following 
sequence of events.  First, the \siivlam\/ absorption begins to weaken 
(indicating a reduction in the \siiv\ column density).  Next, the \nivlam\/  
absorption begins to weaken (indicating a reduction in the \niv\ column 
density).  Finally, the \xmm\/ X-ray flux begins to increase.  

To interpret these results as angular separations on the stellar surface, we 
must remember that the phases refer to the repeating DACs and that the 
stellar rotation period is probably twice as long.  This means that the 
angular separations with respect to the star are half those inferred from 
the phase differences.  Thus, the phase difference between \siiv\ and \niv\ 
translates to $10 \pm 3^\circ$ (\iue) or $23 \pm 9^\circ$ (STIS) and the 
phase difference between \niv\ (STIS) and the X-ray curve becomes $14 \pm 
8^{\circ}$.

%%%%%%%%%%%%%%%%% Analysis of the XMM spectra 
%%%%%%%%%%%%%%%%%%%%%% light curve and softness ratio %%%%%%%%%%%%%%%%%%%%%%
\begin{figure}[t]
\centering
\includegraphics[height=0.9\columnwidth, angle=0]{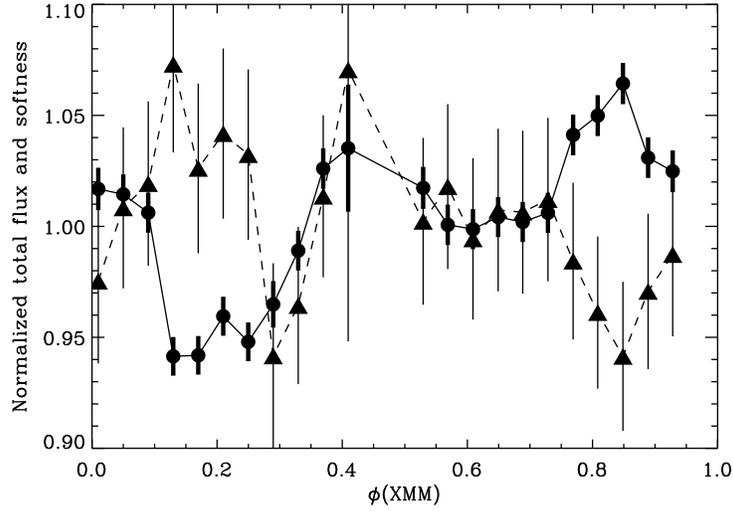}
\vspace{-2.4in}
\caption{Total EPIC flux (circles and solid lines) and softness ratio 
(triangles and dashed lines) plotted against phase.  Both curves are 
normalized by their means.}
\label{fig:soft}
\end{figure}
%%%%%%%%%%%%%%%%%%%%%%%%%%%%%%%%%%%%%%%%%%%%%%%%%%%%%%%%%%%%%%%%%%%%%%%%%%%%

Next, we examine how the spectral properties of the X-rays behave as a 
function of phase.  Figure~\ref{fig:soft} shows the EPIC fluxes as a 
function of phase (normalized by their mean) and the softness ratio defined 
in \S~\ref{sec:xmm} normalized by their mean.  It is clear that the two 
quantities are anti-correlated, with the softness ratio decreasing when the 
X-ray flux increases and vice versa, i.e., the spectrum becomes soft when 
the flux is weak and hard when the flux is strong.  This is at odds with 
expectations.  It has been proposed that X-ray variability in O stars might 
result from additional absorption of X-rays by the material in the CIRs 
\citep[e.g.][]{Osk2001}. In this case, one would expect a harder and more 
strongly absorbed X-ray spectrum during the X-ray minimum, implying that the 
softness ratio should drop during the X-ray minimum.  Instead, 
Fig.\,\ref{fig:soft}, shows that {\em the X-ray emission is softer at 
minimum than at maximum}.  

%%%%%%%%%%%%%%%%%%%%%%%%%%%%%%%%%%%%%%%%%%%%%%%%%%%%%%%%%%%%%%%%%%%%%%%%%%%%
\begin{figure}
\centering
\vspace{.2in}
\includegraphics[width=0.4\columnwidth]{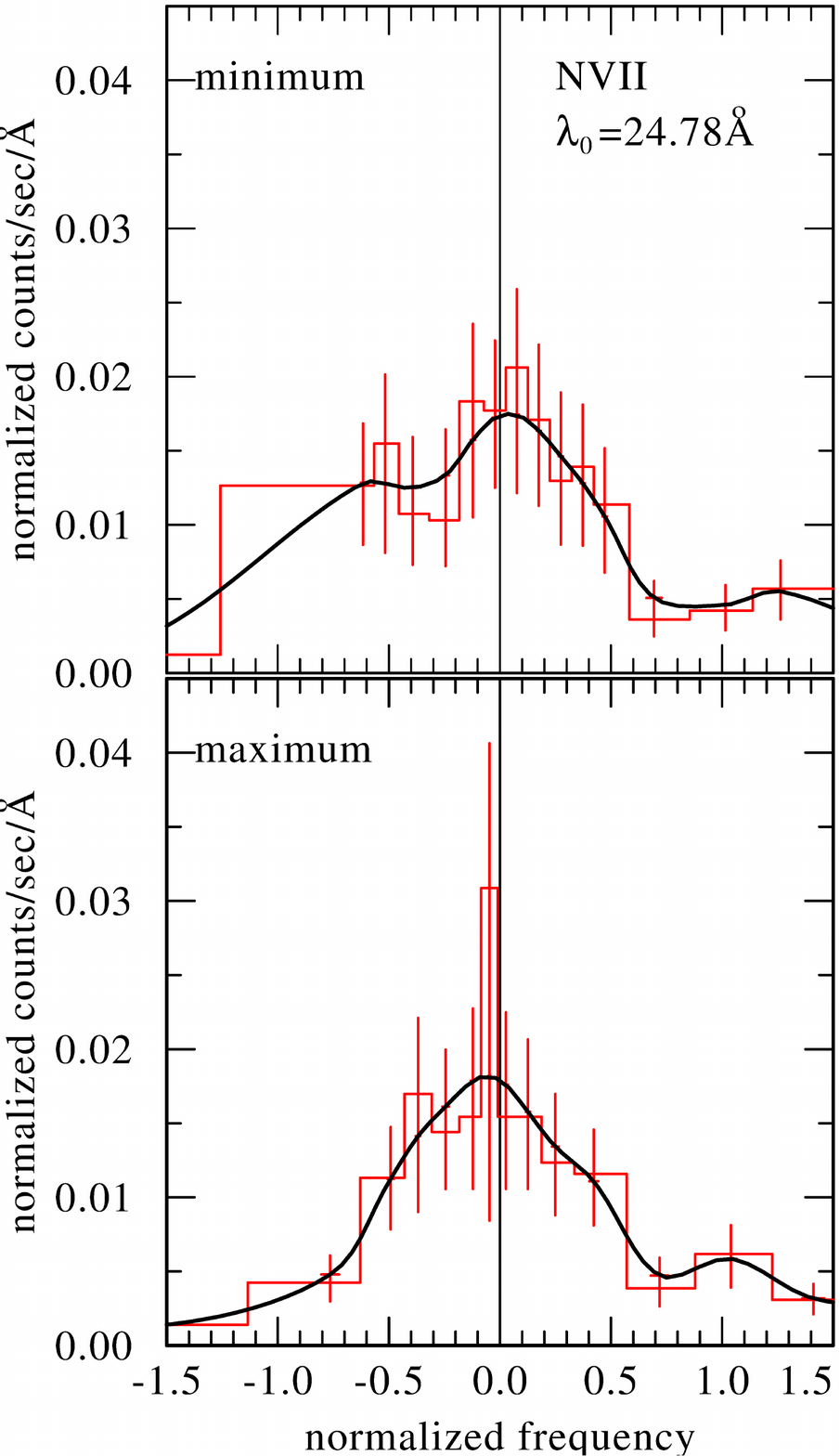}\hspace{-0.8in}
\includegraphics[width=0.4\columnwidth]{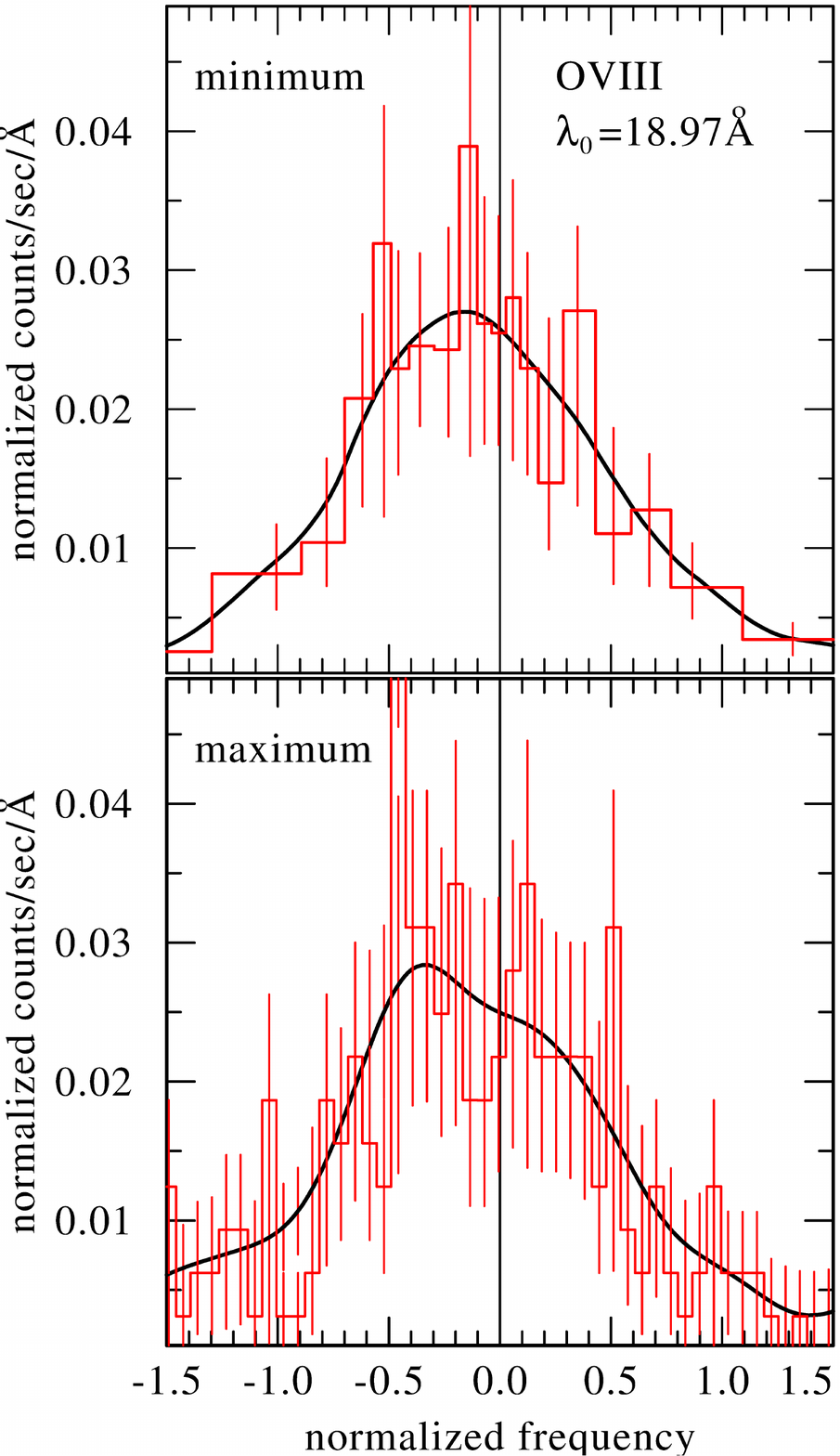}\hspace{-0.8in}% \hfill
\includegraphics[width=0.4\columnwidth]{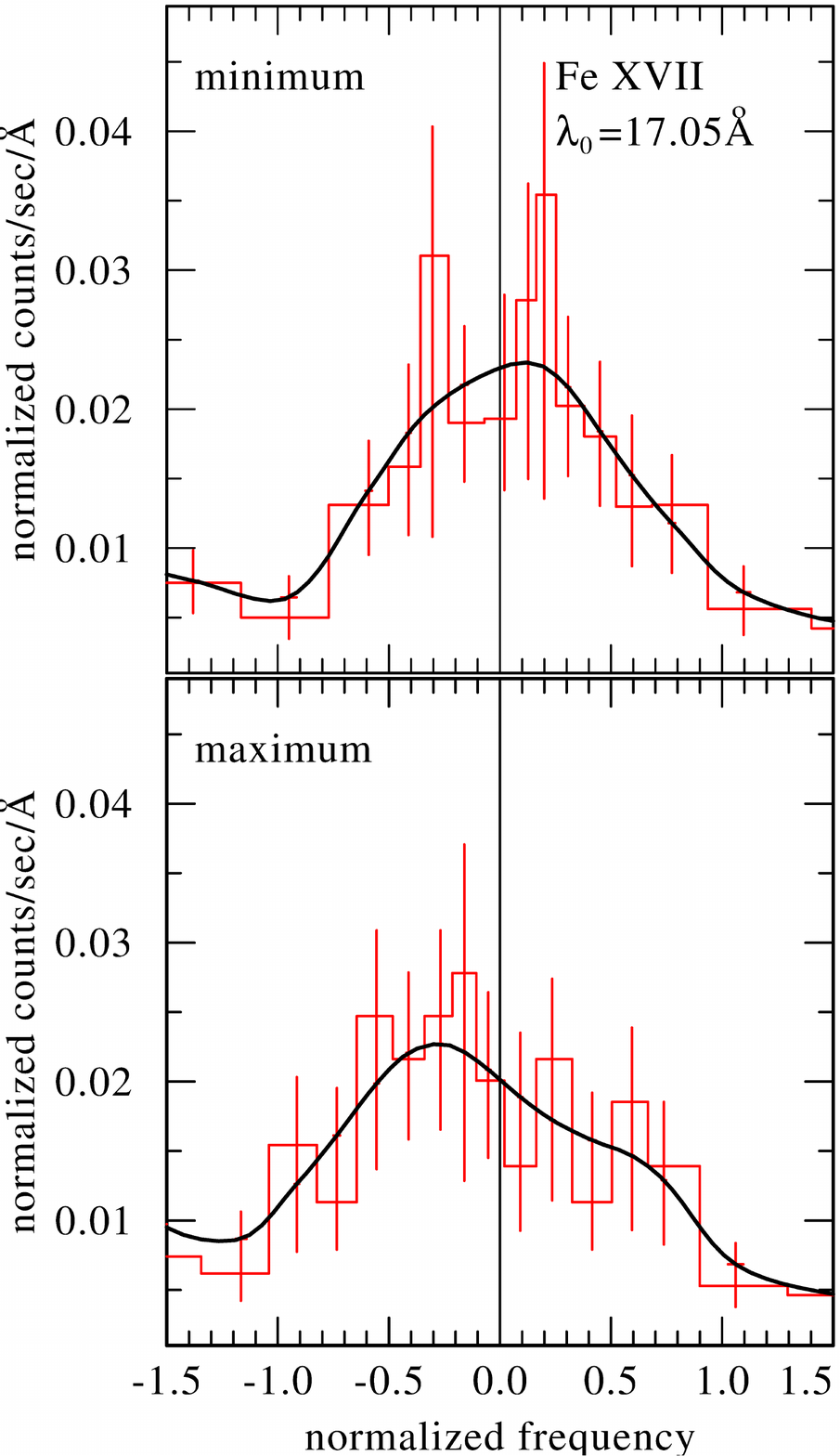}
\caption{Mean profiles for RGS spectra obtained at X-ray minimum (top) and 
maximum (bottom) for N~{\sc vii} (left), O~{\sc viii} (middle) and 
Fe~{\sc xvii} (right) plotted against $v/v_\infty$, where $v_\infty = 2450$ 
\kms.  These lines sample progressively higher ionized gas.  
%A Gaussian fit to the O~{\sc viii} profile is also shown.  
The smooth curves are the data smoothed by a Gaussian with a full width of 
0.08 \AA, in order to accentuate the positions of the curve maxima.
}
\label{fig:xlines}
\end{figure}
%%%%%%%%%%%%%%%%%%%%%%%%%%%%%%%%%%%%%%%%%%%%%%%%%%%%%%%%%%%%%%%%%%%%%%%%%%%%

Finally, we examine the line profiles of the strongest lines in the RGS 
spectra.  For an on-axis source, the accuracy of the RGS first order 
wavelength scale is $0.005$\,\AA.  Figure~\ref{fig:xlines} shows the 
profiles of N~{\sc vii} $\lambda 24.8$, O {\sc viii} $\lambda 18.97$ and Fe 
{\sc xvii} $\lambda 17.1$ whose ionization ranges cover 552 --  667, 739 
-- 871 and 490 -- 1256~ev, respectively.  
Comparison of the smoothed profiles obtained at 
% Comparison of the profiles obtained at 
minimum and maximum reveal an important trend: the strong emission lines 
observed at maximum are more strongly blue-shifted than at the minimum.  
However, this trend is not as evident in N\,{\sc vii} Ly$\alpha$ (discussed 
below), which appears centered on its rest wavelength (taking into account 
line blending) at both minimum and maximum.  \citet{mc1991} and 
\citet{ignace01} showed that emission lines produced by optically thin 
radiation suffering K-shell absorption in stellar winds are expected to be 
skewed and blue-shifted.  Moreover the blue-shift and the line shape is 
sensitive to wind absorption. The higher is the stellar wind opacity, the 
more blue shifted are the emerging emission lines.  With the advent of 
high-resolution X-ray spectroscopy, this formalism was extended to 
accelerating stellar winds \citep{wc2001, oc2001} and clumped stellar winds 
\citep{feld2003}.  Hence, the stronger blue shift of X-ray emission lines 
during X-ray maximum is possibly due to stronger wind opacity at maximum 
than at minimum, but the exact details depend on the geometry of the 
emitting plasma.  Nevertheless, the fact that the N\,{\sc vii}\,$\lambda 
24.78$ line does not appear to shift is consistent with this explanation.  
This line is a blend and its modeling is complicated, but the {\em relative} 
change seen in other lines is not present.  One possible interpretation is 
that its line formation region is different from the heavier ions.  For 
example, \citet{wc2007} pointed out in their analysis of high-resolution 
X-ray spectra of O stars that the line formation regions of heavier ions are 
closer to the stellar surface compared to the lines of lighter ions, again 
suggesting that the increase in X-ray flux is caused by a source deep in the 
wind.      

%%%%%%%%%%%%%%%%%%%%%%%%%%%%%%%%%%%%%%%%%%%%%%%%%%%%%%%%%%%%%%%%%%%%%%%%%%%%
%%%%%%%%%%%%%%%%%%%%%%%%%%%%%%%%%%%%%%%%%%%%%%%%%%%%%%%%%%%%%%%%%%%%%%%%%%%%
\section{Summary and Discussion}\label{sec:disc}
%%%%%%%%%%%%%%%%%%%%%%%%%%%%%%%%%%%%%%%%%%%%%%%%%%%%%%%%%%%%%%%%%%%%%%%%%%%%

The observational results of this paper can be summarized as follows:
\begin{enumerate}
\item The \iue\/ light curves of the low velocity absorption in the UV wind 
lines of \xper\/ vary with a period of 2.086 day and amplitudes of 
$\sim 10$\%.  

\item The periods and shapes of the new STIS light curves are consistent 
with \iue\/ light curves obtained 22 years earlier.

\item Both the \iue\/ and STIS light curves for \siiv\ $\lambda 1402$ and 
\niv\  $\lambda 1718$ are shifted in phase, with the \niv\ curve lagging 
the \siiv\ curve by a phase of $\sim 0.06$.

\item Variations in the X-ray flux are consistent with a 2.086 day period 
and vary with an amplitude of $\sim 10$\%, but have a distinctly different 
curve shape.  Nevertheless, if its minimum is aligned with the minimum in 
the \niv\ line light curve,  it is also shifted in phase, lagging the \niv\/ 
curve by a phase of $\sim 0.08$.  

\item The sequence of events is: \siiv\ maximum, \niv\ maximum, and 
X-ray light maximum. 

\item The X-ray flux is softer at minimum than at maximum, suggesting that 
the X-rays are more strongly absorbed at maximum.

\item The profiles of the X-ray lines appear to become more skewed toward 
high velocity at X-ray maximum.  Furthermore, the lines of heavier ions are 
more skewed at maximum than those of lighter ions.  These observations are 
consistent with the radiation from hotter plasma suffering higher absorption 
at maximum. 

\item While the \xmm\/ observations cover the 2.086 day period quite well, 
none of them sample the same phase more than once.  As a result, we cannot 
firmly conclude that the X-ray light curve repeats.  
\end{enumerate}

%%%%%%%%%%%%%%%%%%%%%%% implications %%%%%%%%%%%%%%%%%%%%%%%%%%%%%%%%%
If we interpret the observations in terms of our spot model, then we can 
infer the following:  
\begin{enumerate} 
\item Both the \iue\/ and STIS light curves are well represented by a simple 
spot model with two identical, bright, surface spots located on opposite 
sides of the star.  Each spot covers roughly 10\% of the projected stellar 
disk, implying radii $R_{spot} \sim 0.3R_\star$ and an angular diameter of 
$\sim 37^\circ$.  These are only representative values, as spots at 
different latitudes would have somewhat different areas.

\item By bright spots, we mean regions of reduced low velocity absorption.  
This could indicate less photospheric absorption or a smaller Sobolev 
optical depth at low velocity.  The latter could be due to a localized 
increase in the wind velocity gradient, reduction in the mass loss rate or 
shift in the ionization state.  
  
\item The similar flux variations in lines with very different ionization 
potentials, oscillator strengths and formation properties suggests that 
their variations are due to variable covering factors and not to varying 
optical depths.   
 
\item Assuming a rotation period of $2 \times 2.086$ day, the phase 
differences between events implies that the \siiv\ minimum occurs first, 
then the \niv\ minimum occurs 23$^\circ$ after that and the X-ray minimum 
occurs $14^\circ$ after that. 

\item The X-ray spectrum at maximum is more absorbed than at minimum, 
which suggests that the source of the variable X-ray flux originates 
deep in the wind.  

\item The X-ray line profiles are more skewed to high velocity at X-ray 
maximum.  The interpretation of this result depends on the geometry of the 
emitting and absorbing regions, making it strongly model dependent.  

\item The fact that the low velocity (near surface) wind line variations 
correlate with the X-rays together with the observation that the X-rays 
are more strongly absorbed at maximum suggests that both variations occur 
very near the stellar surface.
\end{enumerate}

%%%%%%%%%%%%%%%% place below
In addition, we note that whatever the physical mechanism that causes the 
X-ray and UV wind line variations must also cause less than a few 
hundredths of a magnitude in the photometric variability at optical 
wavelengths \citep{Ramiaramanantsoa14}.   Furthermore, this mechanism 
must be quite common, since most hot stars with well developed but 
unsaturated wind lines which have been observed long enough display the 
temporal signature of CIR-like structures.  This includes WRs and CSPNs.  
Although bolstered by fewer examples (see \S \ref{sec:intro}), it is 
becoming clear that whenever an OB star with a strong wind is observed long 
enough in X-rays, variability on a time scale of order the rotation period 
is also revealed.  

Finally, we note that the persistence of two identical, diametrically 
opposed starspots for decades suggests an oblique magnetic rotator, 
where the spots are the magnetic poles.  However, strong limits on the 
existence of even relatively weak dipole fields in \xper\ and similar 
stars \citep{david14} appear to make this mechanism doubtful.

In closing, we point out that a major caveat in our results is that we do 
not have direct evidence that the X-ray light curve repeats.  Although 
our data cover a complete cycle and appear to splice together well, we 
lack observations for the same phase obtained during different cycles.  
Such observations would conclusively prove the connection between the UV 
wind lines and X-rays.  Furthermore, if the two spot model is correct, the 
two segments of our X-ray curve sample the same spot, and information on 
the diametrically opposed spot is totally lacking.  Clearly, more X-ray 
data are needed.  The observations that would cement the connection are 
repeated X-ray observations sampling the same phases of the light curve 
over several periods to determine whether it repeats.  Nevertheless, the 
results of our new multi-wavelength coordinated observing campaign imply 
that the X-ray variability in \xper\/ is correlated with the UV wind line 
variability.  This suggests that the origin of the X-rays in this star, 
and possibly other single O-stars, is linked to the same physical 
mechanisms responsible for the CIRs in stellar winds, which manifest 
themselves as periodic UV wind line variability.

%%%%%%%%%%%%%%%%%%%%%%%%%%%%%%%%%%%%%%%%%%%%%%%%%%%%%%%%%%%%%%%%%%%%%%%%%%%
%%%%%%%%%%%%%%%%%%%%%%%%%%%%%%%%%%%%%%%%%%%%%%%%%%%%%%%%%%%%%%%%%%%%%%%%%%%
\appendix

%%%%%%%%%%%%%%%%%%%%%%%%%%%%%%%%%%%%%%%%%%%%%%%%%%%%%%%%%%%%%%%%%%%%%%%%%%%
This appendix develops a simple model for the light variations we observe 
at low velocity in wind lines, and examines some of its properties.  
We should point out that because we isolate a portion of the profile 
defined by velocity limits, the rotational velocity of the star will cause 
the geometric boundaries of the region to vary as the spot moves around 
the star.  However, this effect is strongest as the spot clears the limb, 
when the projected area of the spot is smallest.  In contrast, the most 
distinctive aspects of the light curves are determined from the spot 
crossing the face of the star, when the projected area is largest and the 
projected rotational velocity smallest.  For similar reasons, limb 
darkening is also ignored.

In modeling the light curves, we must first decide whether the 2.086 day 
period seen in the DACs is the rotational period and the observed 
variability is due to a single spot, or if the rotation period is $2 \times 
2.086$ day and the variability is due to two nearly identical spots 
$180^\circ$ apart.  The observed rotational velocity of \xper\ is 
$v_{obs} = v_{eq} \sin i = 204$ \kms.  Its rotation period, $P$, is 
probably one or two times the DAC period, $P_{obs}$, i.e., $P = P_{obs} 
n = 2.086 n$ day, where $n$ is 1 or 2.  This consideration, along with 
estimates of the probable rotation velocity and stellar radius, places 
limits on $i$.  First, we assume $v_{eq} \lesssim 300$ \kms, otherwise the 
wind line spectrum of \xper\ would probably be abnormal for its spectral 
type \citep[e.g.,][]{massa95, prinja97}.  This gives  $204/300 \leq 
\sin i \leq 1$, or $43^\circ  \leq i \leq 90^\circ$.  Limits on the stellar 
radius restrict $i$ even further since  $R_\star = v_{eq} P/(2 \pi) = 
v_{obs} P_{obs}n/(2 \pi \sin i)  = 8.45 n/\sin i$.  The \citet{weid10} 
tables give a radius of $\sim 14.3$ for an O7 III star.  However, if 
\xper\ is a little less luminous (class III/V), its radius could be as small 
as $R_{min} \simeq 11.8$.  If it is a bit more luminous (class I/III), its 
radius could be as large as $R_{max} \simeq 17.9$.  These imply $8.45\; 
n/R_{max} \leq \sin i \leq 8.45\; n/R_{min}$.  Inserting the limiting radii 
gives $0.47 n \leq \sin i \leq 0.72 n$, or $28^\circ  \leq i \leq 46^\circ$ 
for $n = 1$, and $71^\circ \leq i \leq 90^\circ$ for $n = 2$, where the 
$90^\circ$ limit for $n = 2$ implies $R_{min} = 16.9$.  The $v \sin i$ 
constraint restricts $i$ to $43^\circ \leq i \leq 46^\circ$ for $n=1$, where 
the lower limit infers an $R_{max} = 12.4$.  Together, the constraints give 
the following relations: $43^\circ \leq i \leq 46^\circ$ and $11.8 \leq R_
\star \leq 12.4$ for $n = 1$, and $71^\circ \leq i \leq 90^\circ$ and $16.9 
\leq R_\star \leq 17.9$ for $n = 2$.  Both situations are possible.  The 
inclination, spot size, latitude and intensity of the spots are all free 
parameters, although $i$ is rather strongly constrained for each case.  

In the following, we adopt a two spot model with equatorial spots for the 
simple reason that it is the easiest to calculate and adequately describes 
the light curves.  However, one should keep in mind that opposing spots at 
higher latitudes are possible as are one spot models (if $P = 2.086$ day) 
with the spot at high latitude\footnote{Recently, \citet{gordon18} derive a 
radius for \xper\ of $\sim 11$, which favors one spot models.  However, 
unlike the other stars in their sample, the distance used to determine the 
radius was not from {\it GAIA} data, but from older, marginal {\it 
Hipparcos} data and indirect methods.}. 
%%%%%%%%%%%%%%%%%%%%%%%%%%%%%%%%%%%%%%%%%%%%%%%

Our model consists of two identical, circular, equatorial spots separated by 
180$^\circ$.  These spots can appear either brighter or darker than their 
surroundings, where these terms are referring to wavelengths very near the 
cores of strong lines.  Therefore, a bright spot could be due to a region 
where the strength of the photospheric line is weakened or a region with 
weak low-velocity wind absorption.  Similarly, a dark spot could be due to 
a region where the photospheric line has strengthened or a region with 
strong low velocity wind absorption.  

To quantify the model, let the stellar flux be $f_0$, the flux from the 
spot be $f_s$, and the fraction of the star covered by the spot at 
rotational phase $\psi$ be $a(\psi)$.  There is also a non-variable 
contribution to the flux that comes from the light scattered throughout 
the wind, $f_e$.  With these definitions, the flux varies as
\begin{eqnarray}
f(\psi) & = & [1 -a(\psi)]f_0 + f_s a(\psi) +f_e\\
        & = & (f_0 +f_e) + (f_s-f_0)a(\psi)  
\end{eqnarray}
The size of the spot is defined by the fraction of the disk covered when 
the spot is centered along the line of sight, $a_0 \equiv a(0)$.  In this 
case, the radius of the spot, $R_s$, is $R_s = \sqrt{a_0} R_\star$, where 
$R_\star$ is the stellar radius.  Further, the angular size of the spot 
relative to the center of the star is $\theta_0 = 2\sin^{-1} \sqrt{a_0}$.

To examine the variations of a light curve sampled at $ i = 1, \cdots, N$ 
phases and normalized by its mean, $r(\psi_i) = f(\psi_i)/\langle f(\psi) 
\rangle$,we use a normalized version of the last equation, which is
\begin{equation}
r(\psi_i) = \frac{1 +C a(\psi_i)}{1 +C \langle a(\psi)\rangle} \label{eq:r}
\end{equation}
where  
\begin{equation}
C = \frac{f_s -f_0}{f_0 +f_e} 
\label{eq:const}
\end{equation} 
is independent of phase.  

\begin{figure}
\begin{center}
  \includegraphics[width=0.50\linewidth]{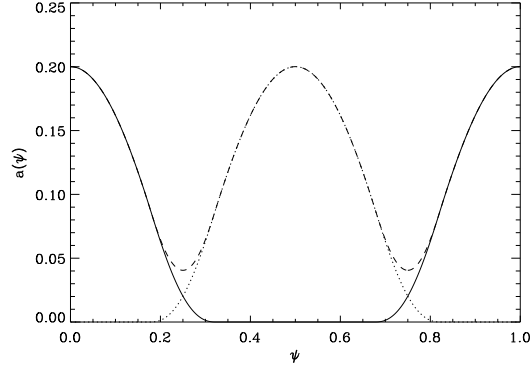}  \vspace{-1.9in}
\end{center} 
\caption{The solid curve is the projected area of a spot whose area is $0.2
\times \pi R_\star^2$ versus phase, $\psi$, where $\psi = 0$ occurs when the 
spot is centered along the line of sight.  The dotted curve is the same 
curve displaced by 0.5 in $\psi$, and represents the projected area of a 
second spot 180$^\circ$ away from the first spot.  The dashed curve is the 
combination of the two and represents the projected area for the 2 spots.} 
\label{fig:areas} 
\end{figure}

The variable component of equation (\ref{eq:r}) is the projected area of 
the spot.  The solid curve in Figure \ref{fig:areas} shows how the projected 
area of a circular, equatorial spot varies as a function of phase for a 
spot with $a_0 = 0.2$.  At $\psi = 0$, the spot is centered along the line 
of sight.  As the star rotates, the area decreases due to projection effects 
until it begins to be occulted by the star.  Once totally eclipsed, the spot 
area remains zero until it begins to emerge from behind the star.  The 
projected area then continues to increase until the spot returns to the 
center of the disk.  The dotted curve is the solid curve displaced in phase 
by 0.5.  When the two are added, the result is the projected area for two 
identical spots on opposite sides of the star.  The combined curve is shown 
as the dashed curve.  The important feature of this curve is that the maxima 
are broader than the minima.  This is simply because the entire spot is 
behind the star for an interval of $\pi -\theta_0$, and some part of the 
spot is visible for an interval of $\pi +\theta_0$.  Note that $\theta_0$ 
can be quite large.  A spot with $a_0 = 0.1$ has an $R_s = 0.32 R_\star$ 
and a $\theta_0 = 37^\circ$.  

\begin{figure}
\begin{center}
  \includegraphics[width=0.45\linewidth]{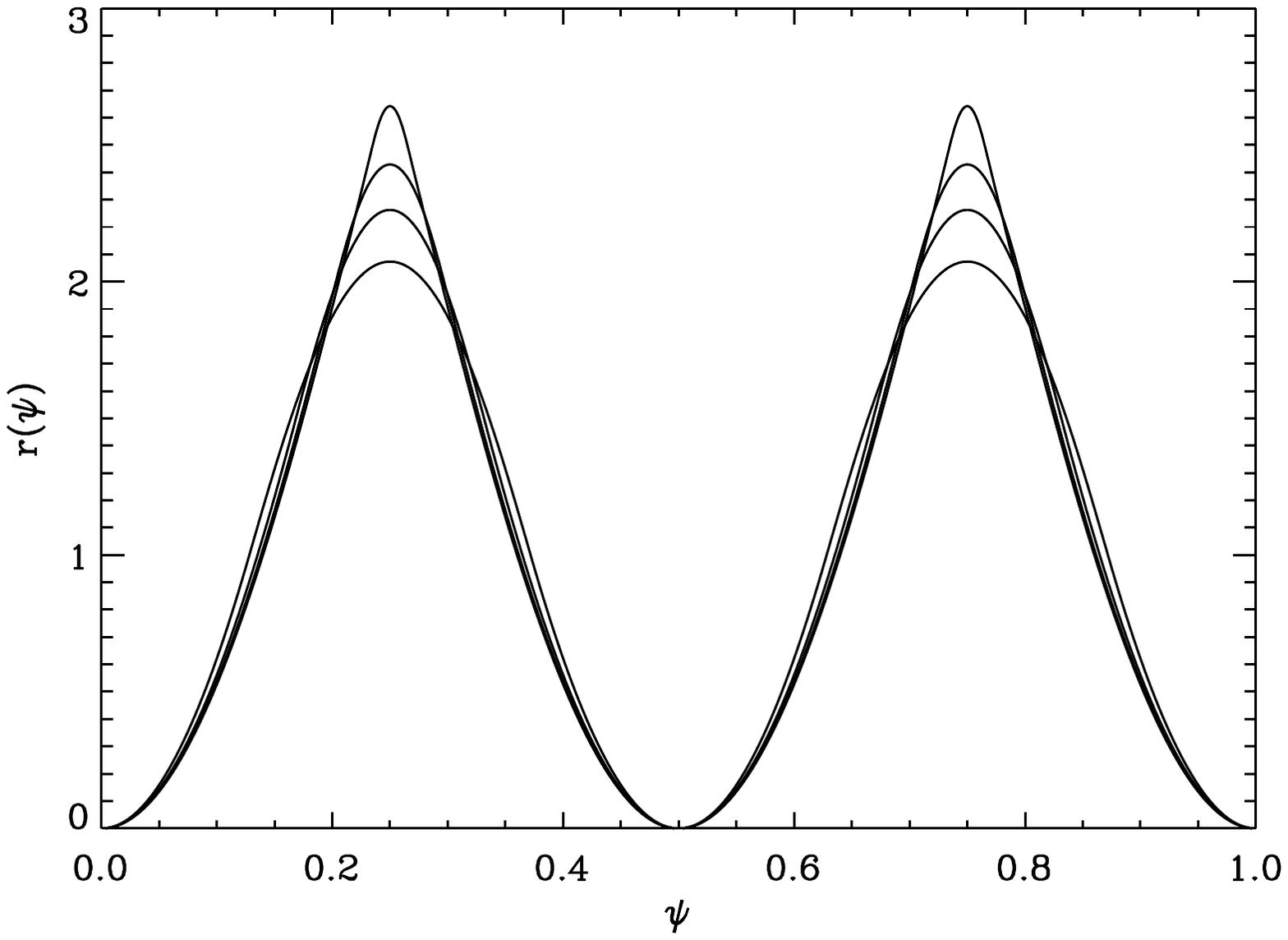}\vspace{-1.0in}
  \includegraphics[width=0.45\linewidth]{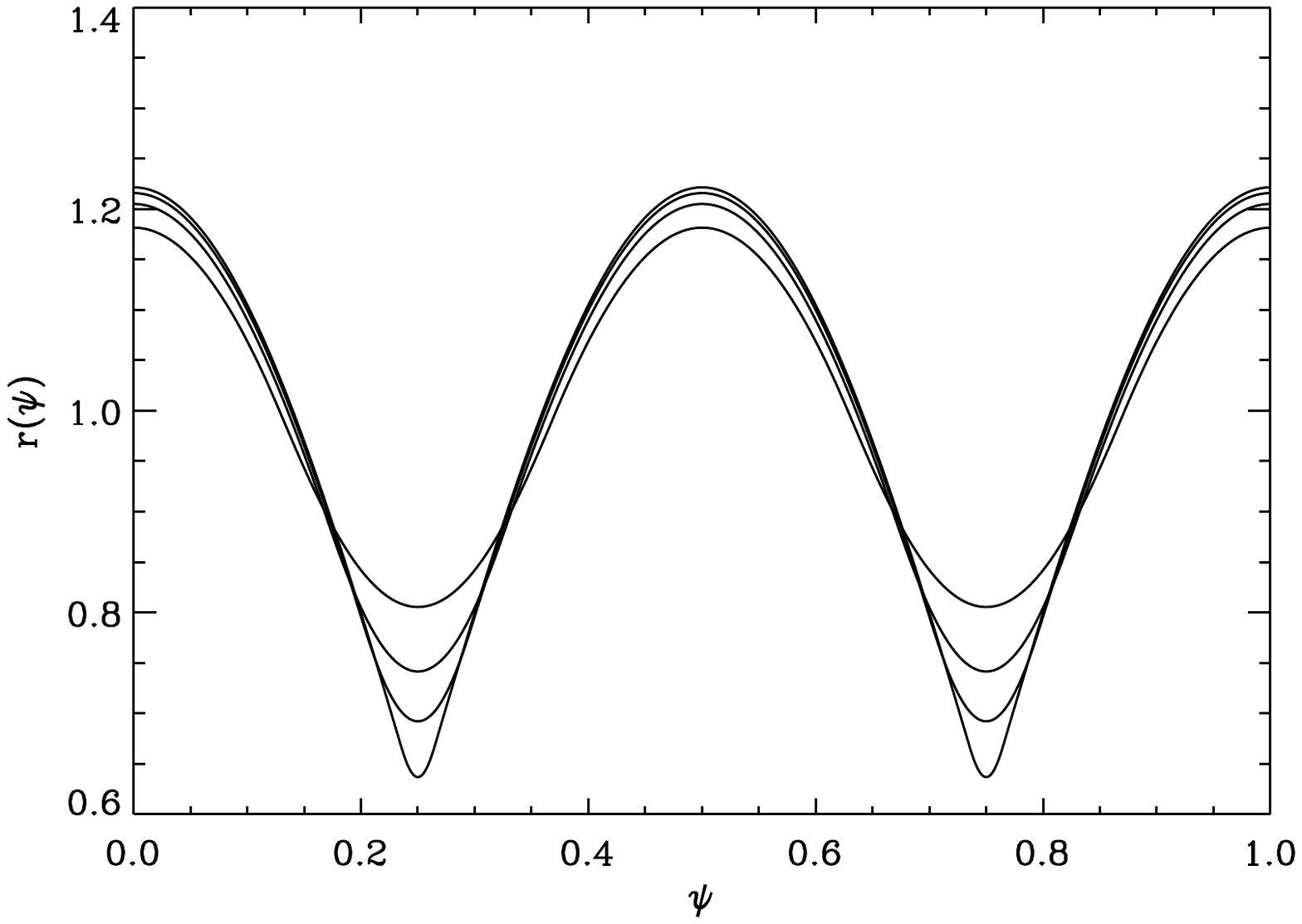}\vspace{-0.7in}
\end{center}
\caption{Left: Model light curves for spots that are optically thicker than 
their surrounding plotted against rotation phase, $\psi$.  The curves are 
for $a_0 = 0.50$, 0.25, 0.10 and 0.01 (bottom to top) and were calculated 
with $|C| = 1/a_0$ to make them similar over most of their range.  Right: 
Similar curves for the case where the spot is optically thinner than its 
surroundings.  In this case curves with smaller $a_0$ have the deepest 
minima.} 
\label{fig:thick_thin} 
\end{figure}

Equation (\ref{eq:r}) shows how the nature of the variability is determined 
by the quantity $f_s -f_0$ in the numerator of $C$.  If $f_s > f_0$ (a 
bright spot), then $C > 0$ and the line varies as $1 +|C| a(\psi)$.  If $C 
< 0$ (a dark spot), the line varies as $1 -|C| a(\psi)$.  Figure 
\ref{fig:thick_thin} shows how the two families of $r(\psi)$ curves respond 
to spot size.  In these figures, we set $C = 1/a_0$, which effectively 
normalizes the curves and emphasizes their differences.  The curves are 
similar near the broad maxima (bright spots) or minima (dark spots), since 
the major factor dictating the projected area is the change in the 
inclination angle to the line of sight, which is independent of spot size.  
The curves differ near the minima (bright spots) or maxima (dark spots), 
which are shaped by how large the spot is and, therefore, how long it is 
partially visible.  

It is important to note that because curves with $C = 1/a_0$ appear so 
similar (Fig.~\ref{fig:thick_thin}), it will be difficult to extract 
independent values of $C$ and $a_0$ from the observations, especially if 
the region of the sharp extrema are not well sampled.  This means that 
errors in the two parameters will be strongly correlated.  As a result, 
changing the values of $C$ and $a_0$ so that $a_0 C$ is constant, will 
have little effect on the quality of the fits.  

%%%%%%%%%%%%%%%%%%%%%%%%%%%%%%%%%%%%%%%%%%%%%%%%%%%%%%%%%%%%%%%%%%%%%%%%%%%%
% made by stopping spotmodel.pro
%%%%%%%%%%%%%%%%%%%%%%%%%%%%%%%%%%%%%%%%%%%%%%%%%%%%%%%%%%%%%%%%%%%%%%%%%%%%
%%%%%%%%%%%%%%%%%%%%%%%%%%%%%%%%%%%%%%%%%%%%%%%%%%%%%%%%%%%%%%%%%%%%%%%%%%%%
% made by drive_seg.pro
%%%%%%%%%%%%%%%%%%%%%%%%%%%%%%%%%%%%%%%%%%%%%%%%%%%%%%%%%%%%%%%%%%%%%%%%%%%%

%%%%%%%%%%%%%%%%%%%%%%%%%%%%%%%%%%%%%%%%%%%%%%%%%%%%%%%%%%%%%%%%%%%%%%%%%%%
In the text, the fluxes are plotted against phase with respect to the 
2.086~day period derived by \citet{dejong01}.  Following de Jong \etal, 
we assume the observed period of the flux variation is half of the rotation 
period, so we introduce $\phi = 2 \psi$, where $\phi$ is the phase of the 
observations relative to the 2.086~day period.  In this case, equation 
(\ref{eq:r}) suggests fitting the observed variations with a model of the 
form
\begin{equation}
r(\phi_i) = \frac{1 +C a(a_0; \phi_i +\phi_0)}{1 +C \langle 
a(a_0; \phi +\phi_0)\rangle} \label{eq:fit}
\end{equation}
where $C$ and $a_0$ are the same as above and $\phi_0$ is the phase shift 
required to align the spot model with the observations.  This is equation 
(\ref{eq:model}) in the text.  

\clearpage
 
\section*{Acknowledgments}
DM and RI acknowlege support Grant \# HST-GO-14180 provided by NASA through 
a grant from the Space Telescope Science Institute, which is operated by 
the Association of Universities for Research in Astronomy, Inc., under NASA 
contract NAS5-26555.  LO is supported by the Deutsches Zentrum f\"ur Luft 
und Raumfahrt (DLR) grants FKZ 15 OR 1809, and partially by the Russian 
Government Program of Competitive Growth of Kazan Federal University.  We 
also thank the referee, whose comments greatly enhanced the clarity and 
conciseness of our presentation.

%%%%%%%%%%%%%%%%%%%%%%%
\bibliographystyle{mn2e}
\bibliography{lo.bib}
\end{document}